\documentclass[prd,a4paper,
nofootinbib,
notitlepage,
longbibliography,
onecolumn,
]{revtex4-1}
\usepackage{graphicx}
\usepackage{amsfonts}
\usepackage{amssymb}
\usepackage{slashed} 
\usepackage{color}
\usepackage{hyperref}
\usepackage{booktabs}
\def\eq#1{{Eq.~(\ref{#1})}}

\def\etal{{\it et al.}}

\definecolor{oucrimsonred}{rgb}{0.6, 0.0, 0.0}
\definecolor{persianblue}{rgb}{0.11, 0.22, 0.73}
\definecolor{forestgreen}{rgb}{0.13,0.35,0.13}
 \hypersetup{colorlinks, citecolor=oucrimsonred, linkcolor=persianblue, urlcolor=oucrimsonred}
\def\hhref#1{\href{http://arxiv.org/abs/#1}{#1}} 
\newcommand{\be}{\begin{equation}}
\newcommand{\ee}{\end{equation}}
\newcommand{\bea}{\begin{eqnarray}}
\newcommand{\eea}{\end{eqnarray}}

\newcommand{\bc}{\begin{center}}
\newcommand{\ec}{\end{center}}
\newcommand{\com}[1]{}
\newcommand{\gsim}{\lower.7ex\hbox{$\;\stackrel{\textstyle>}{\sim}\;$}}
\newcommand{\lsim}{\lower.7ex\hbox{$\;\stackrel{\textstyle<}{\sim}\;$}}

\begin{document}
\title[]{
Mono-chromatic single photon events at the muon collider 
}
\date{\today} 
\author{{ M.\ Casarsa}$^{\dag}$}
\author{{M.\ Fabbrichesi}$^{\dag}$}
\author{{E.\ Gabrielli}$^{\ddag\dag *}$}
\affiliation{$^{\dag}$INFN, Sezione di Trieste, Via  Valerio 2, 34127 Trieste, Italy }
\affiliation{$^{\ddag}$Physics Department, University of Trieste, Strada Costiera 11, 34151 Trieste, Italy}
\affiliation{$^{*}$Laboratory of High-Energy and Computational Physics, NICPB, Ravala 10, 10143 Tallinn, Estonia}
\begin{abstract}
{
\noindent  The cross section for lepton pair annihilation into a photon and a dark photon or an axion-like particle  is constant for large center-of-mass energies because some  of the  portal operators coupling Standard Model and dark sector are proportional to the energy.  Feebly coupled though they are, these  portal operators will be enhanced by the large center-of-mass energy made available by a muon collider and thus provide the ideal example of possible physics beyond the Standard Model to be studied with such a machine.  We discuss  the characteristic signature of the presence of these operators:  mono-chromatic single photon events for the two benchmarks of  having center-of-mass energies of 3 and 10 TeV and  integrated luminosity of, respectively,  1 and 10 ab$^{-1}$. We find that  an   effective scale of the portal  operator as large as  $\Lambda=112$ TeV for an axion-like particle and   $\Lambda=141$ TeV for a dark photon can be separated from the background with a confidence level  of 95\% in the first benchmark; these interaction scales    can be raised to $\Lambda=375$ TeV and $\Lambda=459$ TeV in the case of the second benchmark. The signal for the  pseudo-scalar particle can be distinguished from that of the spin-1 with about 500 events. The response of the detector to high-energy photons is examined.}
 \end{abstract}

\maketitle 

\section{Introduction}

Interest in designing and building a muon collider~\cite{Long:2020wfp,Delahaye:2019omf} will be directly proportional to the capability of  such a collider of exploring not only the Standard Model with great precision but also physics beyond it~\cite{Greco:2016zli,AlAli:2021let}. 
Many a dedicated study  has already been performed, mostly for the physics of the  Higgs boson~\cite{Barger:1995hr} but also  on  precision electroweak physics~\cite{Buttazzo:2020uzc,Han:2020pif}, vector boson fusion processes~\cite{Costantini:2020stv}, lepto-quarks discovery~\cite{Asadi:2021gah}  and possible dark matter candidates~\cite{Han:2020uak}.
 It is in this spirit that we consider the possibility of probing the physics of a dark sector at a muon collider. 

Dark sectors are  made of states that are singlets under the Standard Model gauge groups~\cite{dark_sector}. If they had no other interaction with ordinary matter, they would be invisible but for large-scale gravitational effects, the same way as Dark Matter is. To make these states accessible to our probes,  they are also assumed to have some interaction, dubbed the \textit{portal}, to Standard Model particles; this interaction  is sufficiently feeble to allow for light states, and, indeed, even massless states of the dark sector  to be as yet undetected.
The search for these elusive dark particles has been carried on so far in various  dedicated experiments (see \cite{dark_sector} for a recent review).

Dark particles can couple  to Standard Model states by means of effective higher dimensional operators---starting from and, for all practical purposes, dominated by dimension 5 operators. These  operators appear for instance in   dark-photon coupling to Standard Model fermions via magnetic dipole interactions~\cite{Holdom:1985ag,Fabbrichesi:2020wbt}, as predicted by portal dark-sector models \cite{Dobrescu:2004wz,Gabrielli:2016cut}, or axion-like particles to di-photon couplings, as a consequence of the $U(1)$ Peccei-Quinn anomaly \cite{Peccei:1977hh,Weinberg:1977ma,Wilczek:1977pj}.
On dimensional grounds, the production cross section of a dark particle in association with a photon at high energy  tends to a constant proportional to $1/\Lambda^2$, with $\Lambda$ the effective scale associated to  the dimension five operators. This behavior must be compared  to that of the cross section for dark particles production  by  renormalizable couplings to Standard Model particles, the cross section of which is expected to decrease as $\sigma \sim 1/s$ at high center of mass (CM) energy $\sqrt{s}$. On the other hand, the corresponding cross section for the Standard Model background, characterized by a photon plus a neutrino pair, scales as $1/s$ at high energy \cite{Bento:1985in,Berends:1987zz} which leads to the enhanced ratio of signal over background at high energy for dark-particle productions in association to a photon. 

This feature makes a collider with both  high energy and high luminosity a very promising machine for the study of the dark sector. 
 The muon collider is a case in point. Such a collider can provide the simplest, but also the most striking signature of the existence of a dark sector: the annihilation of the muon pair into a light dark particle and a single photon.
As the dark particle is invisible and assumed to be light, the event is a mono-chromatic single photon with almost half of the center-of-mass energy.

We study the afore-mentioned signature for the two benchmark scenarios of a muon collider with CM energy of 3 and 10 TeV. The   integrated (5 years) luminosity is taken to be of, respectively,  1 and 10 ab$^{-1}$. The signal is enhanced over the background by the large energy though several background events persist in the same energy region because of the radiative return  of  the $Z$-boson pole and a statistical analysis is necessary in order to distinguish the signal from this background.   The mono-chromatic signature at muon collider induced by the radiative return effect for heavy Higgs bosons has been analyzed in \cite{Chakrabarty:2014pja}.

Our analysis shows that within the first five years of operation, the muon collider will be able to set new, and more stringent limits to the effective couplings of dark photon and axion-like particles to muons and photons. The nature of the dark sector particle, whether is a pseudo-scalar coupled to photons or a spin-1 coupled to muons, can also be decided within the first five years of operation. We also discuss in detail the response of the detector to high-energy photons.

Previous work on the dark sector along  similar  lines  includes \cite{Mimasu:2014nea}, where the ALP coupling to photons is studied at hadron colliders, \cite{Bauer:2018uxu,Biswas:2019lcp}, where limits on the same coupling are estimated for the CLIC and FCC-ee colliders, and \cite{Darme:2020sjf}, where the ALP couplings are studied at fixed-target experiments (for instance for the PADME experiment). Our work is the first to discuss the separation from the background of the mono-photon signal and the implementation  to the muon collider. Experimental searches for the same mono-photon signature have been performed at the LEP~\cite{OPAL:1998aqw,L3:2003yon,DELPHI:2003dlq}, the Tevatron~\cite{CDF:2008njt,D0:2008ayi} and the LHC~\cite{ATLAS:2016zxj,CMS:2018ffd} though only providing  rather weak bounds.

\subsection{Dark photons and axion-like particles}

Here we consider two possible candidates for the invisible state in the single photon signature: a massless, spin 1 particle (the dark photon~\cite{Holdom:1985ag,Fabbrichesi:2020wbt}) and a light pseudo-scalar particle~\cite{Weinberg:1977ma,Wilczek:1977pj,Ringwald:2014vqa} (axion-like in its properties).  

The dark photon (DP) $A^\prime _\mu$ with field strength $F^{\prime \mu \nu}$ can couple to the muons via the magnetic-dipole interaction with Pauli dipole term
\be
{\cal L}^{\rm \tiny dipole}_{\rm \tiny DP}= \frac{1}{2\, \Lambda} \, ( \bar \mu \, \sigma_{\mu\nu}\,  \mu  ) \, F^{\prime \mu \nu} \, ,  \label{L-DP}
\ee
where $ \sigma_{\mu\nu}$ is defined to be $i [\gamma^\mu,\, \gamma^\nu]/2$. The scale $\Lambda$ modulates the strength of the interaction.  In a UV completion of the theory the effective scale $\Lambda$ can be generated at one-loop by the exchange of heavy particles in the portal sector \cite{Dobrescu:2004wz,Gabrielli:2016cut}.
The operator in \eq{L-DP} originates from a $SU(2)_L$ invariant  dimension-6 operator of the form
$( \bar L \, \sigma_{\mu\nu}\,  \mu_R ) H F^{\prime \mu \nu}+h.c.$, where $H$  and $L$ are the $SU(2)_L$ Higgs and muon-lepton doublets respectively. Here we consider only its contribution in the broken electro-weak  phase, by replacing the Higgs fields by its vacuum expectation value, as expressed in \eq{L-DP}.

The coupling in \eq{L-DP} is the only one in the case of a massless dark photon. In the massive case,\footnote{
In addition to the  Pauli dipole term, the massive dark photon has also an ordinary coupling to the vectorial muon current
\be
{\cal L}^{\rm \tiny tree}_{\rm \tiny DP}=\varepsilon e ( \bar \mu \, \gamma^{\mu}\,  \mu  ) A^\prime _\mu \, , \label{lead}
\ee
arising from a tree-level contribution of kinetic mixing of dark-photon with ordinary photon \cite{Holdom:1985ag}, that in the massive case cannot be rotated away.} 
the Pauli operator  in \eq{L-DP} has not been constrained by current massive DP searches because all of them have been performed at low-energies ($\sqrt{s}\ll \Lambda$) where its contribution is suppressed by terms of order $s/\Lambda^2$.
It  however becomes relevant at the higher energies of a muon collider.
The effective interaction approach is consistent  as long as the effective scale $\Lambda$ is  assumed to be larger or at most of the same order as the CM energy.

The axion-like particle (ALP) $a$ couples to the muons by means of the portal operator
\be
{\cal L}^{\rm \tiny muon}_{\rm \tiny ALP}=\frac{1}{ \Lambda}  \, ( \bar \mu \, \gamma_5   \gamma^\mu \, \mu  ) \, \partial_\mu a \label{L-ALP1}
\ee
and to photons by means of
\be
   {\cal L}^{\rm \tiny photon}_{\rm \tiny ALP}=\frac{1}{\Lambda} \,  a \, F^{\alpha\beta} \widetilde F_{\alpha \beta} \, , \label{L-ALP2}
\ee
where $\widetilde F_{\alpha \beta}=1/2\epsilon_{\alpha\beta\mu\nu}F^{\mu\nu}$ is the dual field strength of the photon, with $\epsilon_{\alpha\beta\mu\nu}$ the Levi-Civita antisymmetric tensor satisfying  $\epsilon_{0123}=1$. 
For on-shell ALP production the interaction in \eq{L-ALP1} is equivalent to the renormalizable interaction $ (m_{\mu}/\Lambda) \, ( \bar \mu \, \gamma_5  \, \mu  ) \, a$, whose coupling is  proportional to the muon mass $m_{\mu}$ over the $\Lambda$ scale and therefore chirally suppressed.
We restrict our analysis to the ALP production via the effective interaction in \eq{L-ALP2}. The effective field theory expansion is in this case consistent as long as $\sqrt{s} < \Lambda$.

To make contact with the notation of the experiment searches for these interactions, we notice that the  coupling  in the dipole operator in \eq{L-DP} is usually expressed in the literature as the (dimensionful) coefficient $g_{a\mu} = 2/\Lambda$ for the experiments probing the interaction between ALP and muons, which can be identified with ours for the DP because of the similar structure of the effective operators in \eq{L-DP} and \eq{L-ALP1}.
In a similar manner, the ALP-photons interaction in \eq{L-ALP2} is usually expressed in the literature by means of the coefficient  $g_{a\gamma}=4/ \Lambda$.

\subsection{Constraints}

 \begin{figure*}[h!]
\begin{center}
\includegraphics[width=5in]{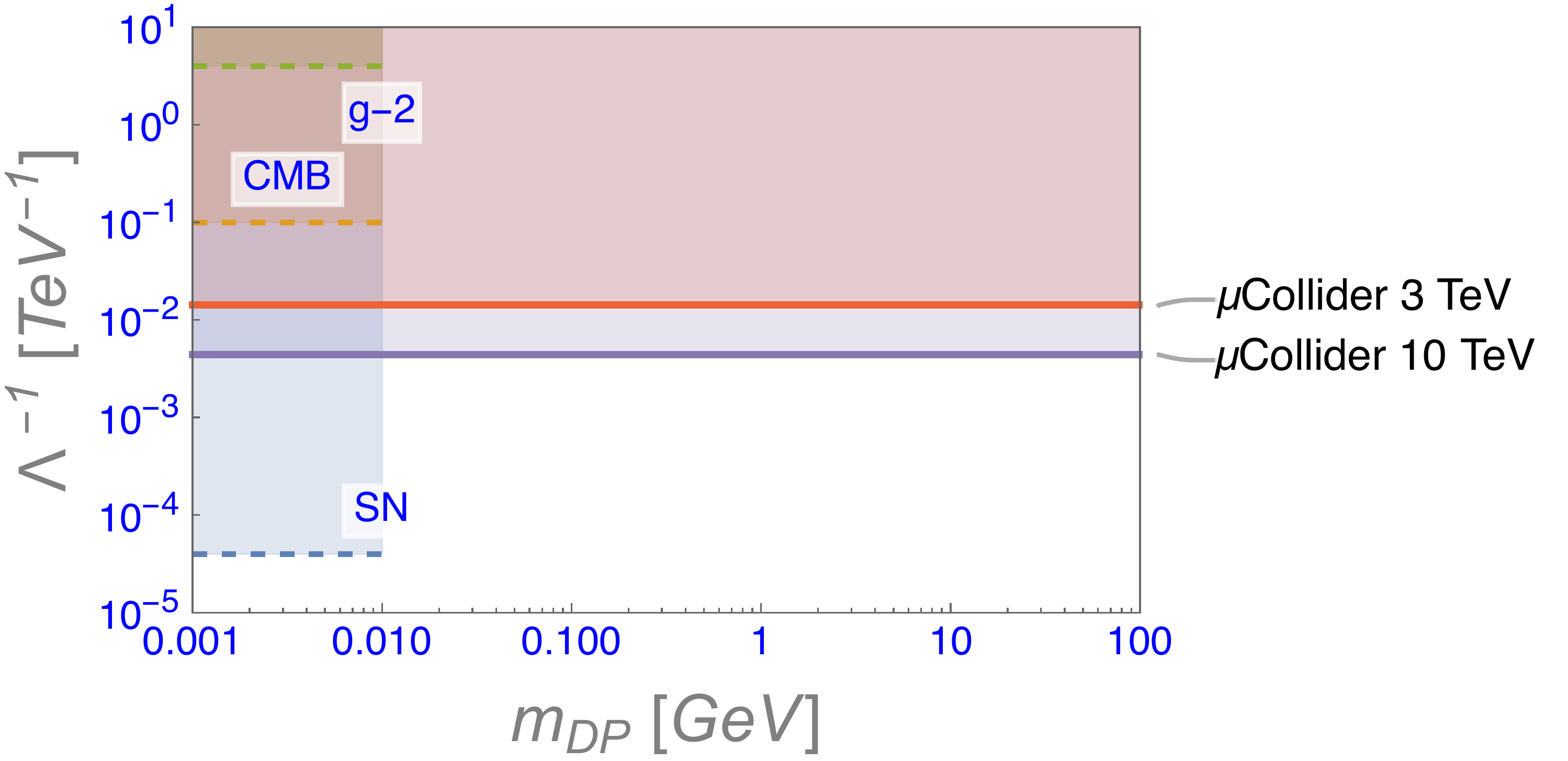}
\caption{\small 
  \label{bounds1}  Limits on DP effective coupling $\Lambda$ to muons as defined in \eq{L-DP}, as a function of the dark-photon mass $m_{A^{\prime}}$: for SN the scale of the coupling to muons has been set at  $10^{4.4}$ TeV  \cite{Bollig:2020xdr} by the effect of dark radiation on  Supernovae dynamics.  For CMB see~\cite{DEramo:2018vss}. For $g-2$ see~\cite{Pospelov:2008zw,Escudero:2019gzq}.   For masses up to 100 GeV the $\mu$Collider limits are for all practical purposes mass independent. 
}
\end{center}
\end{figure*}

 The possibility of seeing DP or ALP at a collider experiment depends on the size of the effective scale $\Lambda$ controlling their interaction with ordinary matter and photons. For the interaction between muons and DP, this scale is mostly constrained by the value of the anomalous magnetic moment $(g-2)$, the number of relativistic species in the early Universe (CMB) and the energy emission of Supernova 1987A (SN). Figure~\ref{bounds1} summarizes these three constraints for the  DP in terms of the effective coupling $1/\Lambda$  and gives the relative references.
 
The  very strong SN bound makes impossible even for a muon collider to explore this  interaction. For this reason, we mostly assume that the dark sector particles are massive with a mass of at least  10 MeV, an energy  scale that makes their production in cosmological and astrophysical process suppressed. 

 \begin{figure*}[h!]
\begin{center}
\includegraphics[width=5.5in]{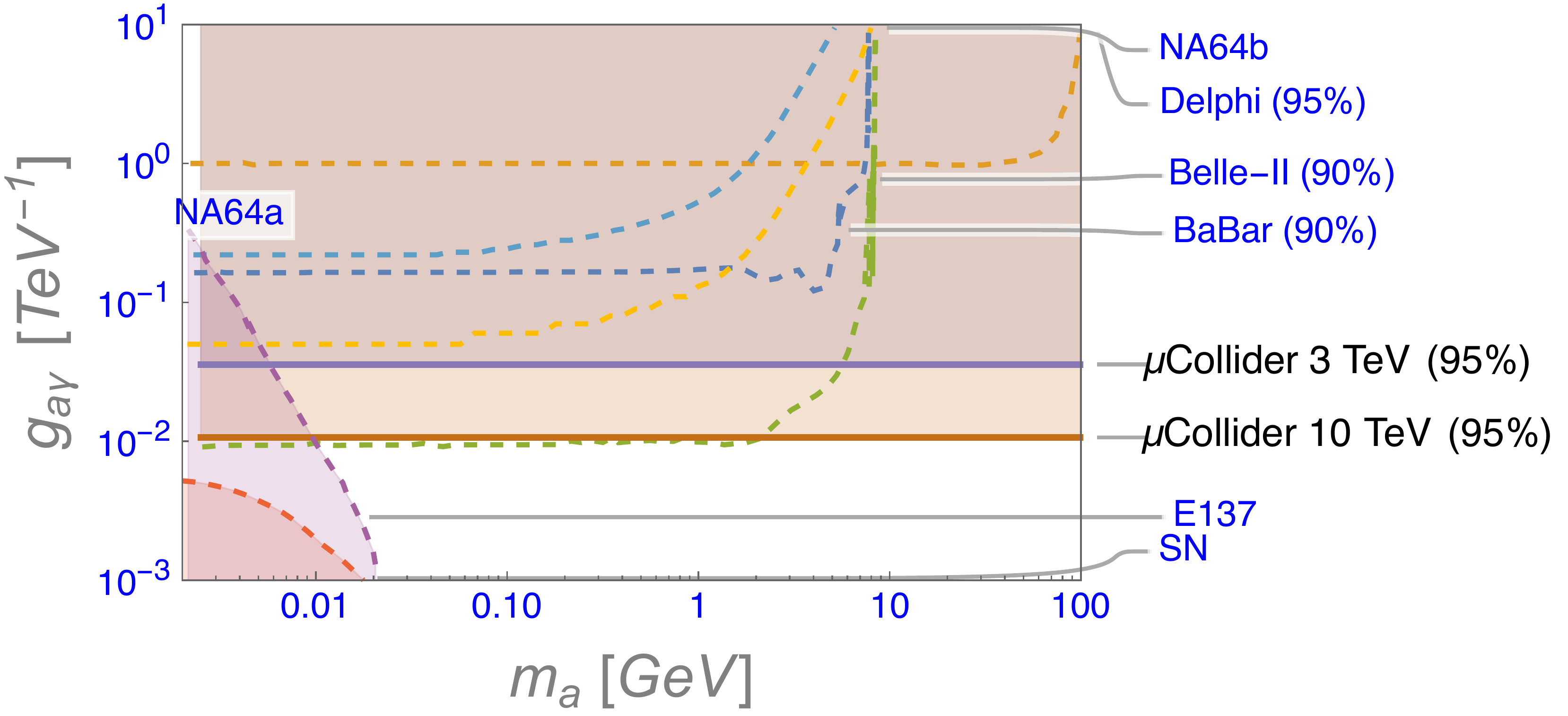}
\caption{\small 
\label{bounds2} Limits on $g_{a\gamma}=4/ \Lambda$ as a function of the ALP mass $m_a$: NA64a~\cite{NA64:2020qwq}, Delphi~\cite{DELPHI:2008uka}  and BaBar~\cite{BaBar:2017tiz} are actual limits. Belle-II~\cite{Belle-II:2018jsg,Dolan:2017osp}, NA64b~\cite{NA64:2020qwq} and $\mu$Collider [this paper] are future estimates. The limit indicated by E137 is  the one from \cite{Bjorken:1988as} as modified for a small ($10^{-4}$) visible branching fraction~\cite{Darme:2020sjf}. For masses up to 100 GeV the $\mu$Collider limits are for all practical purposes mass independent. }
\end{center}
\end{figure*}

The coupling between ALP and  photons is  constrained by direct searches and from SN1987A data. In this case, however, the limit stops for stronger couplings leaving values $10^2 < \Lambda/{\rm GeV} < 10^5 $ possible (again, for a mass larger than 10 MeV).  Figure~\ref{bounds2} summarizes the current limits in terms of $g_{a\gamma}=4/ \Lambda$   and shows future bounds, those in this article included.

We assume that DP and ALP decays are dominated by those into dark sector particles so that their signature remains as missing energy. For this reason, bounds like those from beam dump and other experiments, coming from the decay into visible states are not included.
We  comment below on what range of masses and coupling are consistent with a non-negligible branching rate into Standard Model states.

\begin{figure}
\bc
\includegraphics[width=3in]{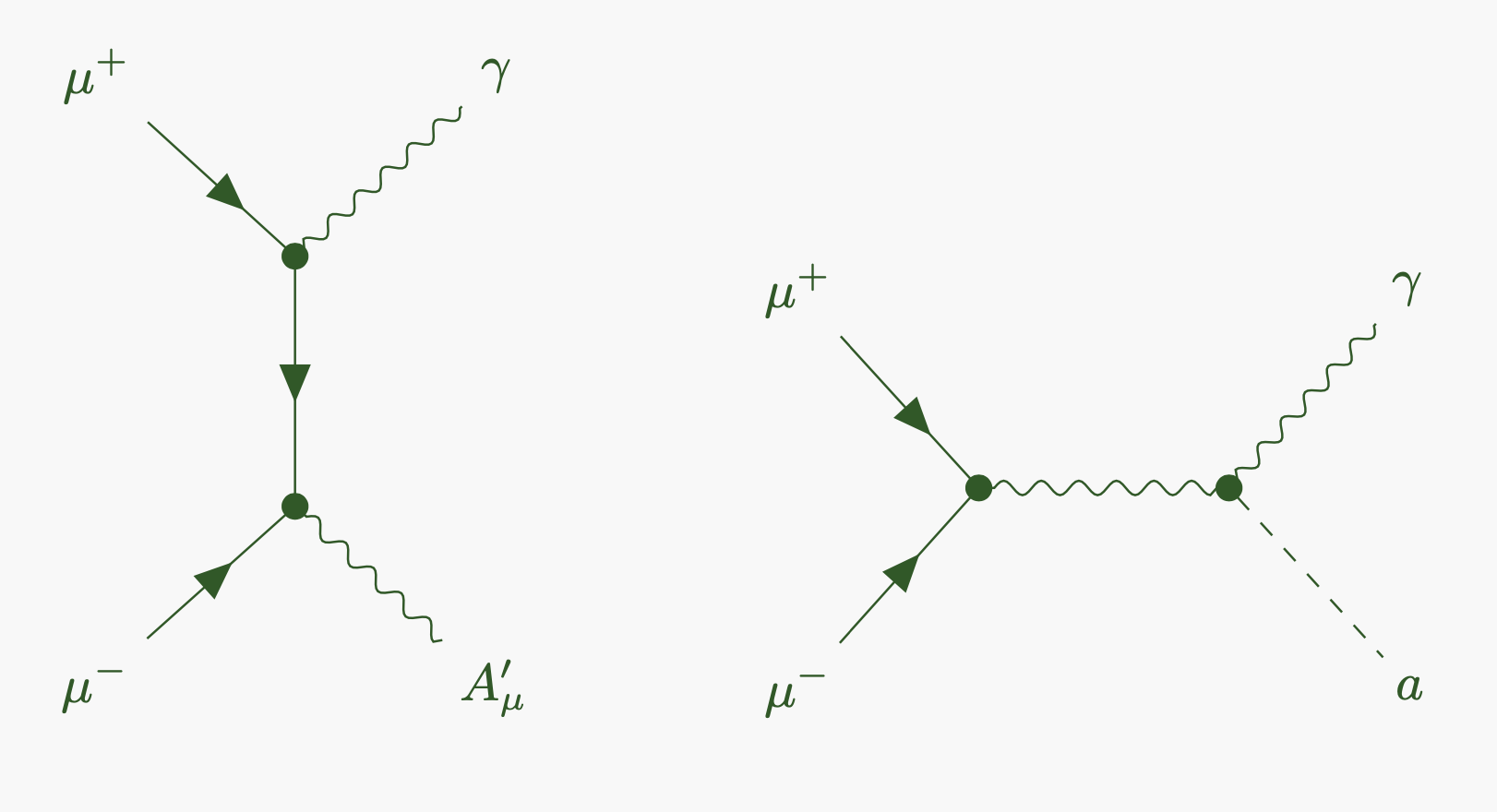}
 \ec
\caption{\small The  diagrams contributing to the processes $\mu^+\mu^-\to \gamma A_\mu^\prime$ (\underline{left}) and $\mu^+\mu^-\to \gamma a$ (\underline{right})  in the high-energy regime. For masses up to 100 GeV the $\mu$Collider limits are mass independent. \label{fig:diagrams}}
\end{figure}

The relevant Feynman diagrams are shown in Fig.~\ref{fig:diagrams}. These two are the only diagrams contributing in the high-energy regime because the direct coupling of the ALP to muons is not enhanced by going to large CM energies.

\subsection{The background: $\mu^+ \mu^- \to \gamma \nu \bar \nu$}

 \begin{figure*}[ht!]
\begin{center}
\includegraphics[width=2.8in]{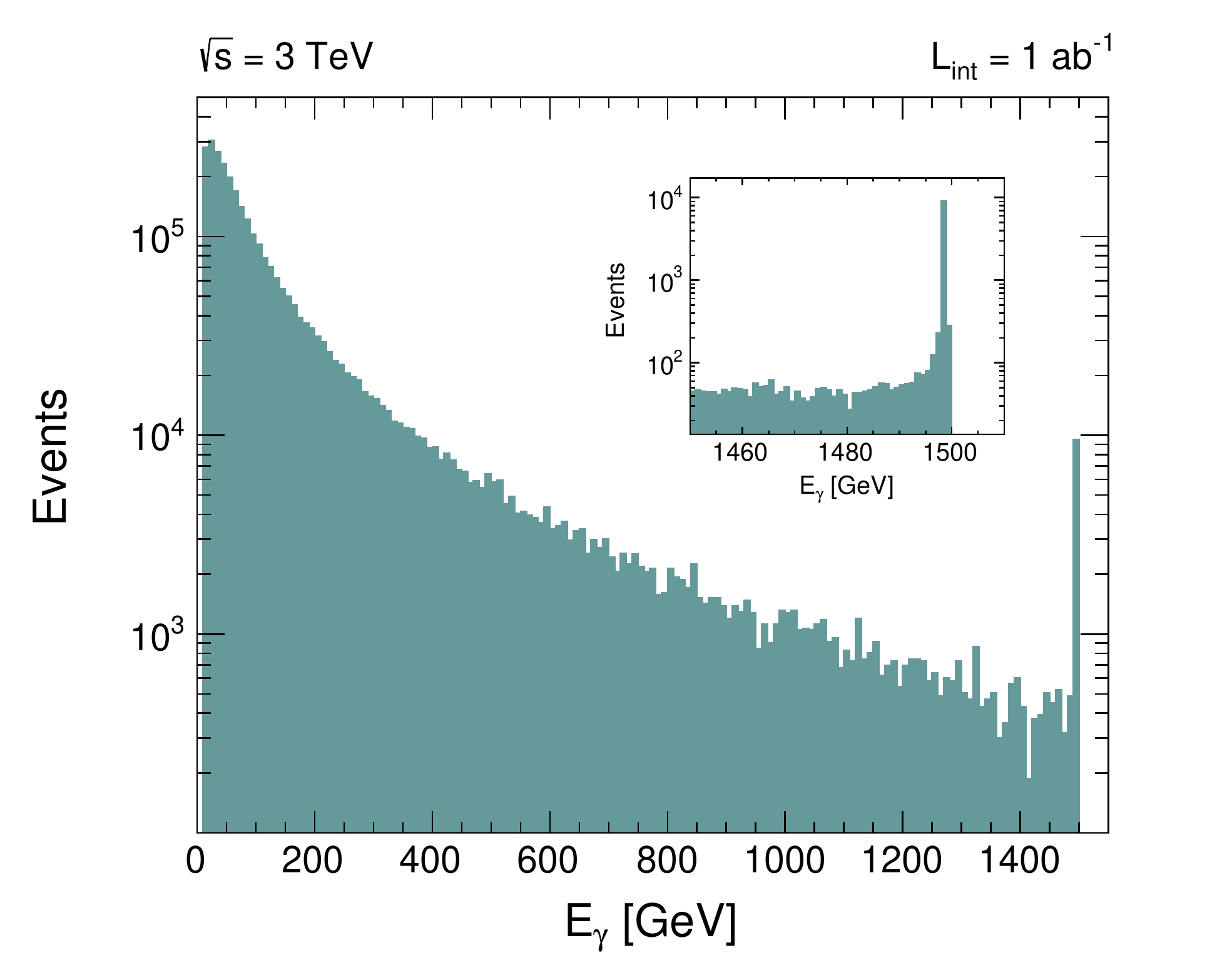}
\includegraphics[width=2.8in]{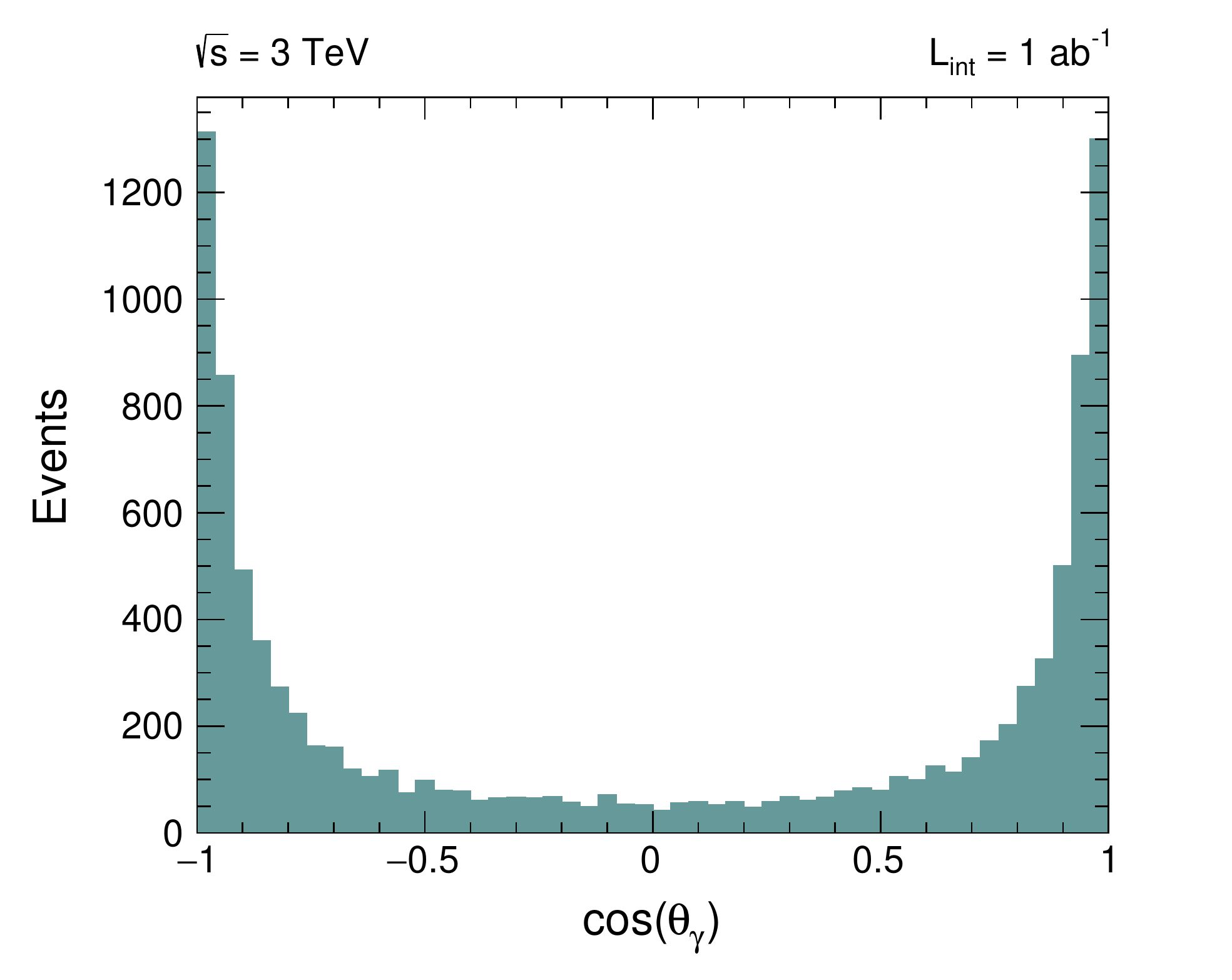}
\caption{\small
\label{fig:bkg}  Background process $\mu^+ \mu^- \to \gamma \nu \bar \nu$: the distributions in $E_\gamma$ and $\cos(\theta_\gamma)$ (the latter for energies larger than 1490 GeV) of the mono-photons. The inset in the left panel shows the tail of the energy distribution. Notice the peak at the end of the energy spectrum due to the radiative return of the $Z$-boson pole.}
\end{center}
\end{figure*}

The Standard Model  process $\mu^+ \mu^- \to \gamma \nu \bar \nu$ gives rise to the same signature as the signal we are after. The analytical expression for the double differential cross section in the solid angle and photon energy is provided in \cite{Bento:1985in} and in the low energy regime in \cite{Berends:1987zz}.  Figure~\ref{fig:bkg} shows the $E_\gamma$ and $\cos(\theta_\gamma)$ distributions for the background events.
The cross section grows with the CM energy but the number of events with a high-energy photon  decreases. Events at the end of the photon energy spectrum  around
\be
E_\gamma=\frac{\sqrt{s}}{2} \left( 1 - \frac{m_Z^2}{s} \right)
\ee
are enhanced by the radiative return of the $Z$-boson pole.
This feature (unfortunate, for our analysis) reduces the sensitivity to the signal 
that---it being a two-body process---is centered in the same range of energies (for  $s\gg m_Z^2$).

\subsection{Decay length}

 \begin{figure*}[ht!]
\begin{center}
\includegraphics[width=2.6in]{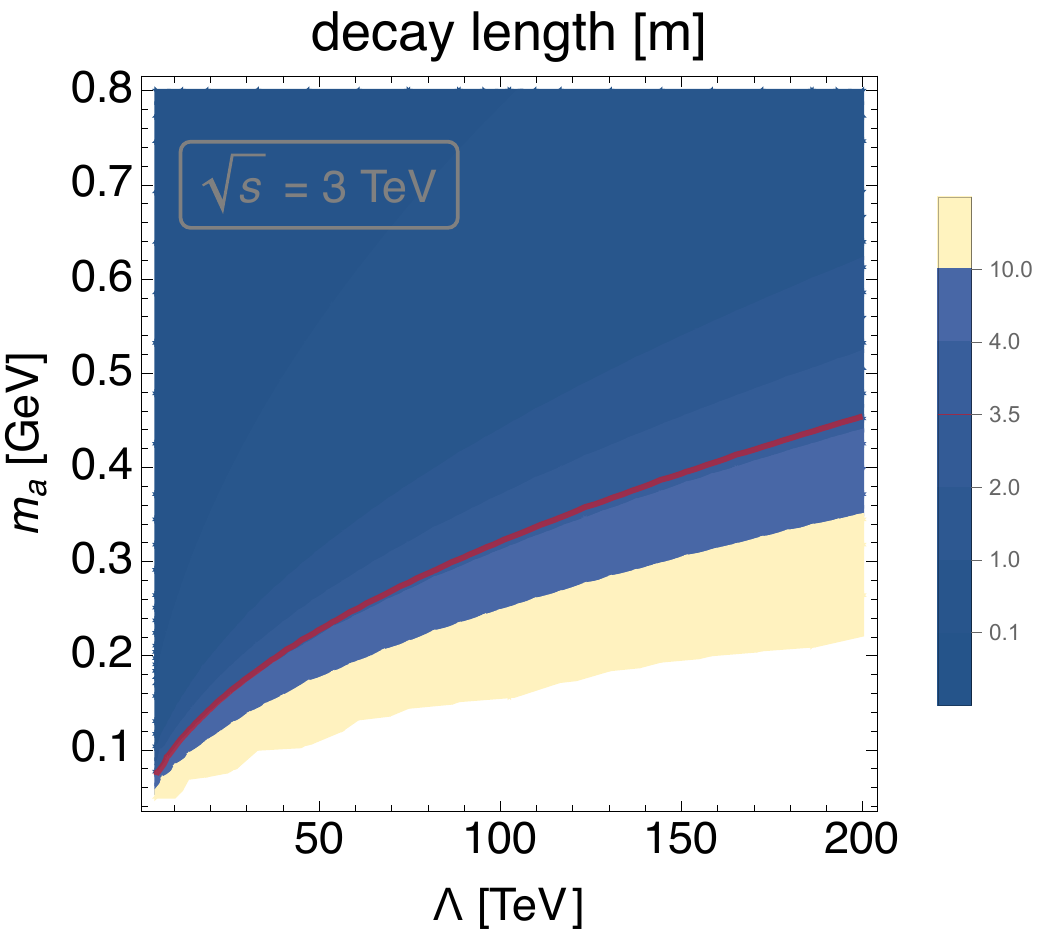}
\includegraphics[width=2.6in]{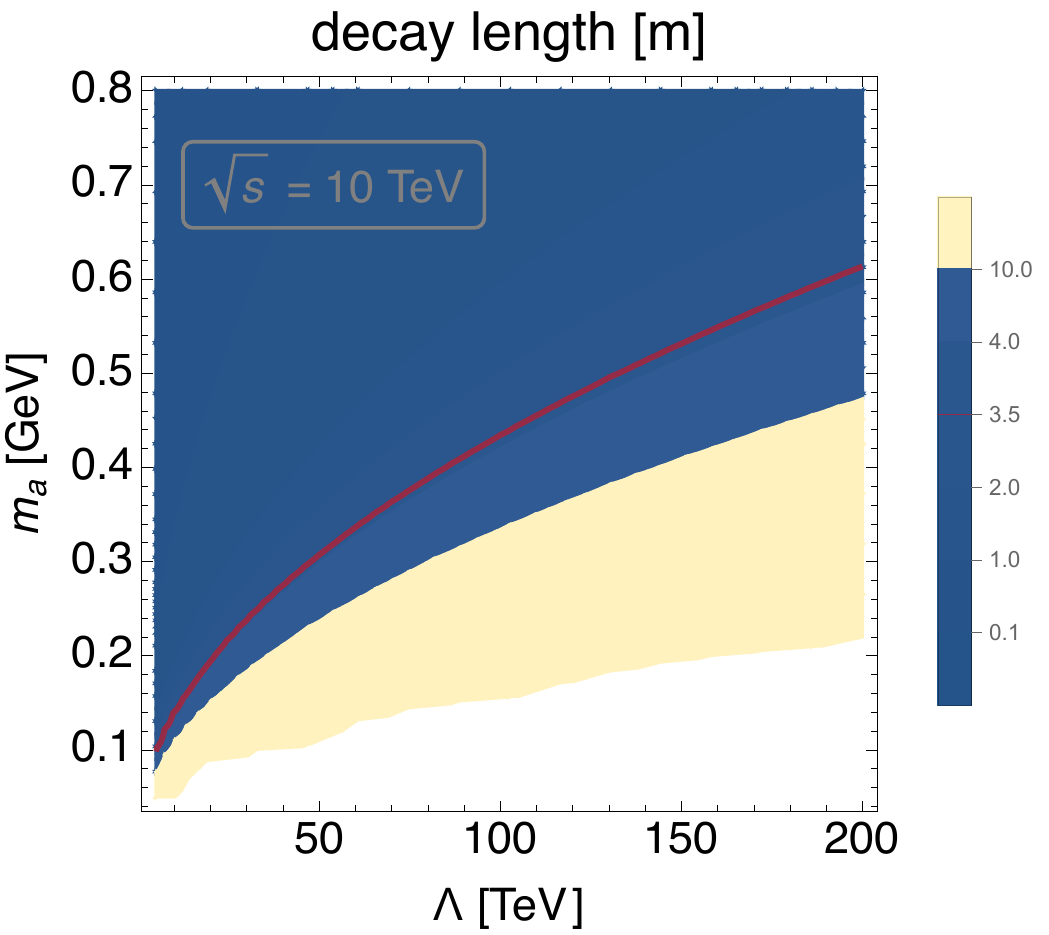}
\caption{\small 
\label{dL-ALP} Decay length (in meters) for the ALP as a function of its mass $m_a$ and the interaction scale $\Lambda$. The contour  in red marks the  size of the detector (3.5 meter). }
\end{center}
\end{figure*}

 \begin{figure*}[ht!]
\begin{center}
\includegraphics[width=2.6in]{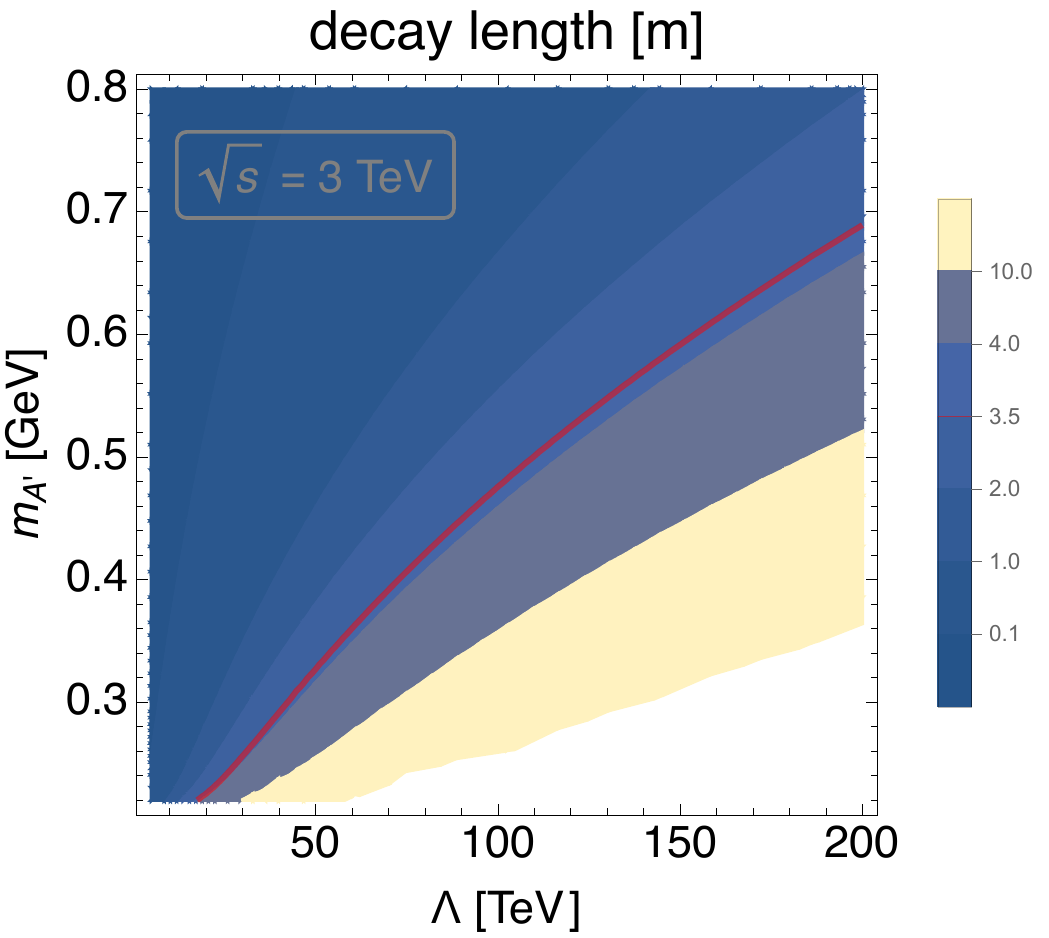}
\includegraphics[width=2.6in]{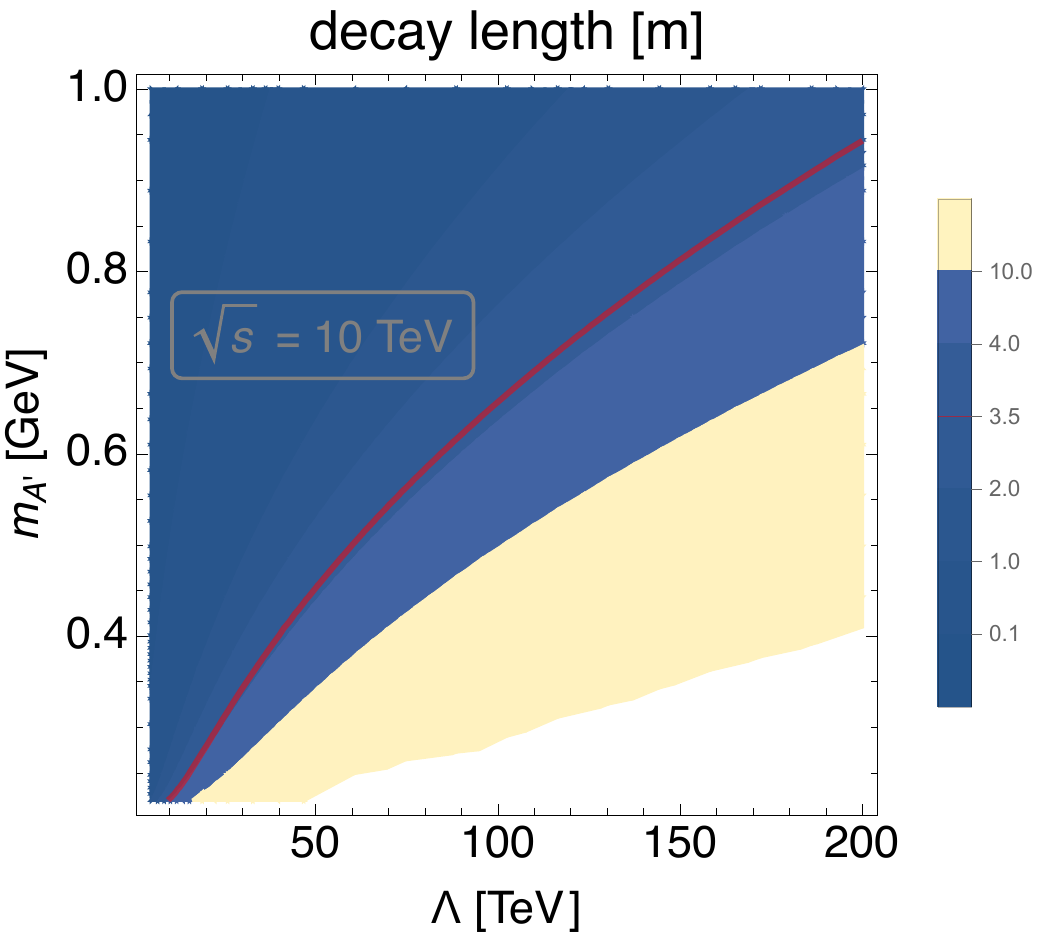}
\caption{\small 
\label{dL-DP} Same as in Fig.~\ref{dL-ALP} for the DP.}
\end{center}
\end{figure*}

If the decay into Standard Model states is non-negligible (or  the dominant one), the decay length 
\be
L =  |\vec{\beta}| \tau=\frac{\sqrt{s}} {2\, m\, \Gamma } \left( 1 - \frac{m^2}{s}\right)\,,
\ee
for a dark state of mass $m$ and width $\Gamma$, with $\sqrt{s}$ the muon  CM energy, may turn out to be inside the detector turning the invisible into a visible track.

Assuming the minimal requirement of a massive dark-photon coupled via magnetic-dipole type of operator to muons, decaying into two muons and assuming axion decaying mainly into two photons, we have for the corresponding decay widths the following results
\bea
\Gamma(A^\prime \to \mu^+\mu^-)&=&\frac{m_{A^\prime}^3(1-4r)^{1/2}(1+8r)}{24\pi\Lambda^2}\, ,\nonumber \\
\Gamma(a \to \gamma\gamma)&=&\frac{m_a^3}{4\pi\Lambda^2}\, ,
\eea
where $r=m_{\mu}^2/m_{A^\prime}^2$, with $m_{\mu}$, $m_A^\prime$, and $m_a$ the masses of the muon, dark photon and axion respectively.  The assumption of a single decay channel  provides an upper bound on the decay length consideration.
In particular, the requirement that the state decay outside the decay length $L$ gives the following lower bound of $\Lambda$ versus the corresponding boson mass $m_{A^\prime}$ or $m_a$,
\bea
\Lambda &>& m_{A^\prime}^2 \frac{L^{1/2}}{s^{1/4}}\frac{(1-4r)^{1/4}(1+8r)^{1/2}}{
  2 \sqrt{3 \pi} (1-m_{A^\prime}^2/s)^{1/2}}~~~\mbox{for DPs}\, ,\nonumber\\
\nonumber\\
\Lambda &>& m_a^2 \frac{L^{1/2}}{s^{1/4}}\frac{1}{\sqrt{2 \pi} (1-m_a^2/s)^{1/2}}~~~\mbox{for ALPs}\, .
\eea

We show in  Figs.~\ref{dL-ALP} and \ref{dL-DP}  the contour plots of the decay length  as a function of  the DP and ALP masses and the corresponding scale $\Lambda$, assuming only the interactions in \eq{L-DP} and \eq{L-ALP2} for the dark-photon and axion respectively.
For masses below approximately 300 MeV (3 TeV) and 400 MeV (10 TeV), and for  energy scales in the coupling as those we explore (which are around and above 100 TeV), the decay length lays outside the detector (taken here to be about 3.5 meter wide) for the ALP.  The same occurs for masses below approximately 500 MeV (3 TeV) and 600 MeV (10 TeV) for the DP.
For masses above  these ranges  the decay may occur inside the detector. For such values we  assume the DP and ALP to decay mostly into dark states and remain invisible.

\section{Methods} 

\subsection{Muon collider} 

Among the projects currently under study for the  generation of particle accelerators following the LHC, the
muon collider represents a unique machine that has the capability to provide leptonic collisions in
a multi-TeV energy range~\cite{Long:2020wfp}.
Such a great physics potential is accompanied by unprecedented technological challenges 
on the experimental side, due to the fact that muons are unstable particles. The electrons
and positrons, created in muon decays, and the photons radiated by them interact with the machine
elements and produce an intense flux of $O(10^{10})$ secondary and tertiary particles 
(photons, neutrons, electrons and positrons, charged hadrons, muons) that eventually may reach the 
detector. 
The amount and the characteristics of the beam-induced background in the detector depend on the 
collider energy and the machine optics and lattice elements. The main features of beam-induced 
background particles are relatively soft momenta (a few MeV for the electromagnetic component, 
half a GeV for the hadronic component, and a few tens GeV for muons) and asynchronous arrival 
times to the detector with respect to the collisions~\cite{BIB}. 

The exploitation of the full physical potential that a muon collider can offer will
depend on the capacity of the experiment to mitigate and cope with the beam-induced background
through cutting-edge technologies and a dedicated design of the machine-detector interface 
and the detector (optimized geometry, high granularity, timing information), new sophisticated 
algorithms for pattern recognition and reconstruction of physical objects~\cite{MuColDet1, MuColDet2}.
In this study, we assume that this is the case and the physical objects we are using are not
significantly affected by the beam-induced background.

The  International Muon Collider Collaboration~\cite{Long:2020wfp} is currently focusing on two muon 
collider conceptual designs: a 3-TeV collider providing an instantaneous luminosity of a few $10^{34}$ cm$^{-2}$ 
s$^{-1}$ and a machine at $\sqrt{s} = 10$ TeV or above with an instantaneous  luminosity of a 
few $10^{35}$ cm$^{-2}$ s$^{-1}$. 
Accordingly, in our analysis, we consider two benchmark scenarios:
\begin{itemize}
\item[\bf S1:]  CM energy of 3 TeV and  total integrated luminosity of 1 ab$^{-1}$, 
\item[\bf S2:]  CM energy of 10 TeV and total integrated luminosity of 10 ab$^{-1}$,
\end{itemize}
and study the generation of events with a single, monochromatic photon plus missing energy in the final states.

\subsection{Event generation and detector simulation}

The events for the signal and the background are generated by means of \textsc{MadGraph5}~\cite{MG5}. 
A 10-GeV cut on the photon generated transverse momentum is imposed to remove most of the soft radiation. The output of \textsc{MadGraph5} is automatically fed into \textsc{Pythia}~\cite{Sjostrand:2006za,Sjostrand:2014zea} and the events thus generated are processed by the detector simulation.

The simulation of the detector is one of the crucial steps in making the study of physics at the muon collider possible. An agreed upon standard is yet to be defined. We utilize detector full simulation tools based on CLIC's ILCSoft framework~\cite{ILCSoft}.
The detector model is based on CLIC's detector concept~\cite{CLICdet},
which comprises a full-silicon tracking system, hermetic high-granularity
electromagnetic and hadronic calorimeters, and a muon spectrometer. 
The tracking detectors and the calorimeters are immersed in a 3.57-Tesla solenoidal magnetic field.
The vertex detector and the machine-detector interface have been adapted to cope with the harsher background environment at a muon collider: the geometry of the vertex detector is optimized to minimize the occupancy from beam-induced particles and two tungsten conical shields, placed around the beam pipe inside the detector, reduce
the background levels in the detector by approximately three orders of magnitude.

The sub-detectors used in this analysis are the electromagnetic (ECAL) and the hadronic (HCAL) calorimeters.
The ECAL and HCAL are both sampling calorimeters.
The ECAL consists of 40 layers of 1.9-mm tungsten absorber and 5$\times$5-mm$^{2}$ 
silicon pad sensors for a total of 22 radiation lengths, 
whereas the HCAL has 60 layers of 19-mm steel absorber and 30$\times$30-mm$^2$ plastic 
scintillating tiles, which correspond to 7.5 nuclear interaction lengths.

\subsection{Event reconstruction and selection}

The full-simulated events are reconstructed with a particle-flow algorithm~\cite{Pandora}, which is integrated in the ILCSoft reconstruction software.
The photon signature in the detector is represented by an isolated electromagnetic shower in the ECAL, to which no charged particle trajectory is associated.
The photon four-momentum is determined from the energy measured in the ECAL and the photon line of flight, taken as the direction from the primary interaction vertex to the position of the energy deposit.

A sample with a single photon per event was used to determine and tune the detector performance in reconstructing and
identifying high-energy photons. The photons, generated in the nominal collision vertex at the center of the detector, are uniformly distributed in energy between 1 GeV and 5 TeV, in polar angle between $10^\circ$ and $170^\circ$, and in the full azimuthal angle range. The photon sample was processed with the detector full simulation and reconstructed with the same algorithms as those used for the signal and background samples. 

It results that a few percent of the energy released by high-energy photons in the ECAL spills into the HCAL. Therefore, for a more accurate reconstruction of such photons, the HCAL energy is recovered and added to the energy of the closest spatially compatible ECAL deposit.
In the same sample, the photon reconstruction performance is assessed:
photons with $E_\gamma \gtrsim 10$ GeV are reconstructed with an efficiency close to 100\%, 
while the photon relative energy resolution is lower than 0.7\% for $E_\gamma \gtrsim 1500$ GeV.
Moreover, corrections are calculated and applied to the photon reconstructed energy to make the detector response uniform as a function of photon energy and polar angle.

Events are selected for the analysis if only one photon has been reconstructed inside the 
detector angular acceptance $10^\circ < \theta_\gamma < 170^\circ$.
The signal is strengthened against  the background by two simple  kinematical cuts:
\begin{itemize}
\item photon energy:  $E_\gamma > 1450$ GeV for $\sqrt{s} = 3$ TeV and $E_\gamma > 4800$ GeV for $\sqrt{s} = 10$ TeV;
\item photon polar angle: $40.4^\circ < \theta_\gamma < 139.6^\circ$;
\end{itemize}
which  optimize the selection process of the events around the end of energy spectrum where the signal is.
In the 3-TeV analysis, this selection yields 3020 background events. The DP signal events range between 193 and 
104 as $\Lambda$ increases from 110 to 150 TeV, whereas the ALP signal events range between 184 and 59 as $\Lambda$ 
increases from 85 to 130 TeV.
In the 10-TeV analysis, this selection yields 2630 background events. The DP signal events range between 160 and 96 
as $\Lambda$ increases from 380 to 490 TeV; the ALP signal events range between 174 and 77 as $\Lambda$ increases 
from 280 to 420 TeV.

  Beside the physical backgrounds from SM processes, we have investigated an additional potential source 
of experimental backgrounds that might mimic the mono-photon signature we are looking 
for.
Using a sample of $\mu^+\mu^-\to\gamma\gamma$ events at $\sqrt{s} = 3$ TeV, we studied 
the case of two-photon events, in which one of the photons is not reconstructed,
and found such a background negligible.

For completeness, we also considered a potential background from events with an energetic 
electron misidentified as a photon. The electron misidentification rate was estimated in a sample of single electrons, processed through the detector full simulation and reconstruction. It is found to be lower than 1\% in the central region of the detector, making also this background negligible.

\section{Results} 

To assess the reach of the muon collider to the considered dark-sector signals,
we test the signal strength for different values of $\Lambda$ against the background-only hypothesis.
The observable that is found to better discriminate the signal from background events is the photon transverse momentum:
the distributions in $p_T^\gamma$ of the signal and the background differ significantly, with the Standard Model background presenting a long tail toward smaller momenta. These two distributions are used to define a likelihood for the background-only (${\cal L}_{\text{B}}$) and the background plus signal hypotheses (${\cal L}_{\text{S+B}}$).

 \begin{figure*}[ht!]
\begin{center}
\includegraphics[width=3in]{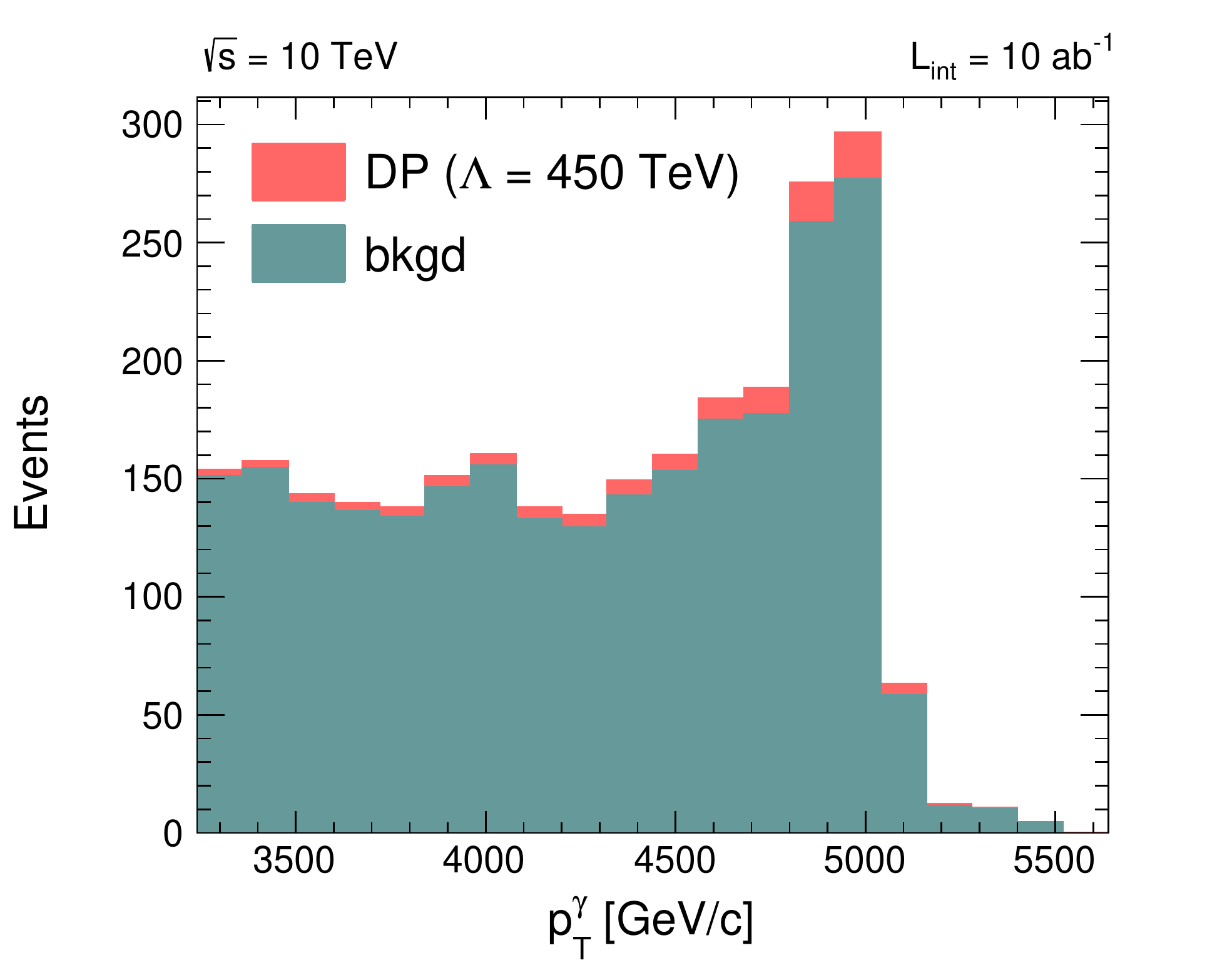}
\includegraphics[width=2.5in]{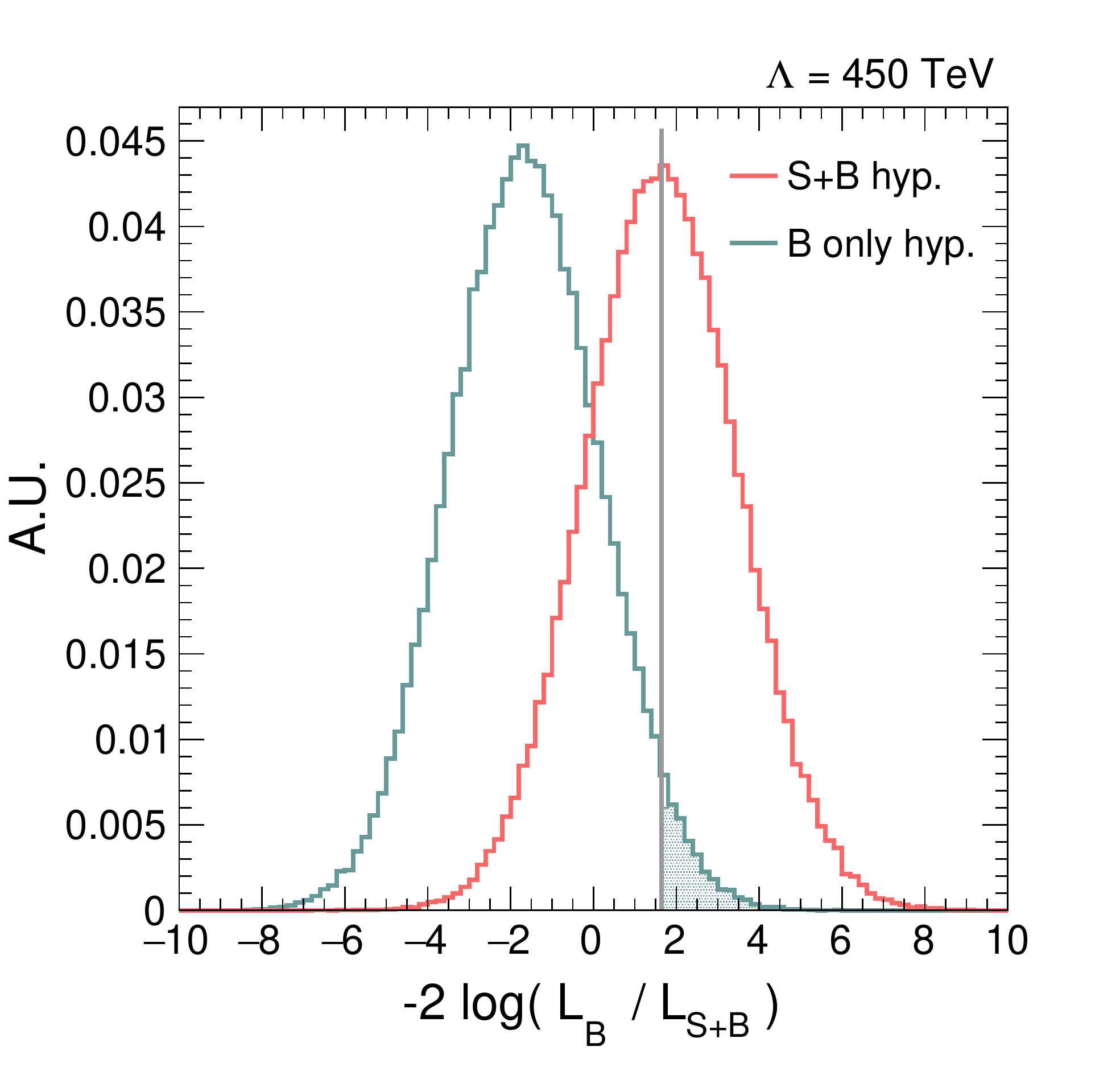}
\caption{\small 
\label{LLR}  Distribution of the photon transverse momenta at $\sqrt{s} = 10$ TeV (left). A
dark photon signal, corresponding to $\Lambda = 450$ TeV, is stacked on top of the background distribution.
The $p_T^\gamma$'s reconstructed above 5000 GeV are due to detector resolution effects and a remnant of the energy correction procedure. 
In the right panel, the LLR distributions  for the background-only and backgrouns-plus-signal hypotheses are shown. 
}
\end{center}
\end{figure*}

The probability distribution functions (pdf) for the null hypothesis (background only) and the alternative hypothesis (background plus signal) can be extracted directly from the dependence on $p_T^\gamma$.  
The signal pdf (pdf$_{\text{S}}$) is parameterized with a crystal-ball function centered at $\sqrt{s}/2$,
while the background pdf (pdf$_{\text{B}}$) is modeled with a complementary error function plus a Gaussian.

We take $N_{\text{obs}}^{(S)}$  signal events as well as $N_{\text{obs}}^{(\text{B})}$ background events generated according to the corresponding pdf's. 
Each event $i$ is characterized by the value of  $p_T^\gamma$. The  likelihood function, for example for the signal plus background hypothesis,    is given by

\be
\label{eq:likelihood}
{\cal L}_{\text{S+B}}  =  e^{-N_{\text{obs}}^{(S)}- N_{\text{obs}}^{(\text{B})}} 
\times \prod_i^{N_{\text{obs}} } \left[ N_{\text{obs}}^{(S)}  \times \text{pdf}_{\text{S}} (p^T_i)  +N_{\text{obs}}^{(\text{B})} 
\times \text{pdf}_{\text{B}} (p^T_i)  
\right] \, ,
\ee
where the events $p^T_i$ are taken from the  $S$ and $B$ populations. In this way, it is possible to randomly generate $N_{\text{obs}}$ events and compute the logarithm of the likelihood ratio (LLR) defined by
\be
\text{LLR} =-2\log \frac{{\cal L}_{\text{B}} }{{\cal L}_{\text{S+B}}}\, .
\ee
By repeating this pseudo-experiment $N_{\text{ps}}$ times, we construct a sample that can be used to compute the LLR statistical distribution for the two hypotheses. We take $N_{\text{ps}}=10^5$.

 \begin{figure*}[h!]
\begin{center}
\includegraphics[width=2.6in]{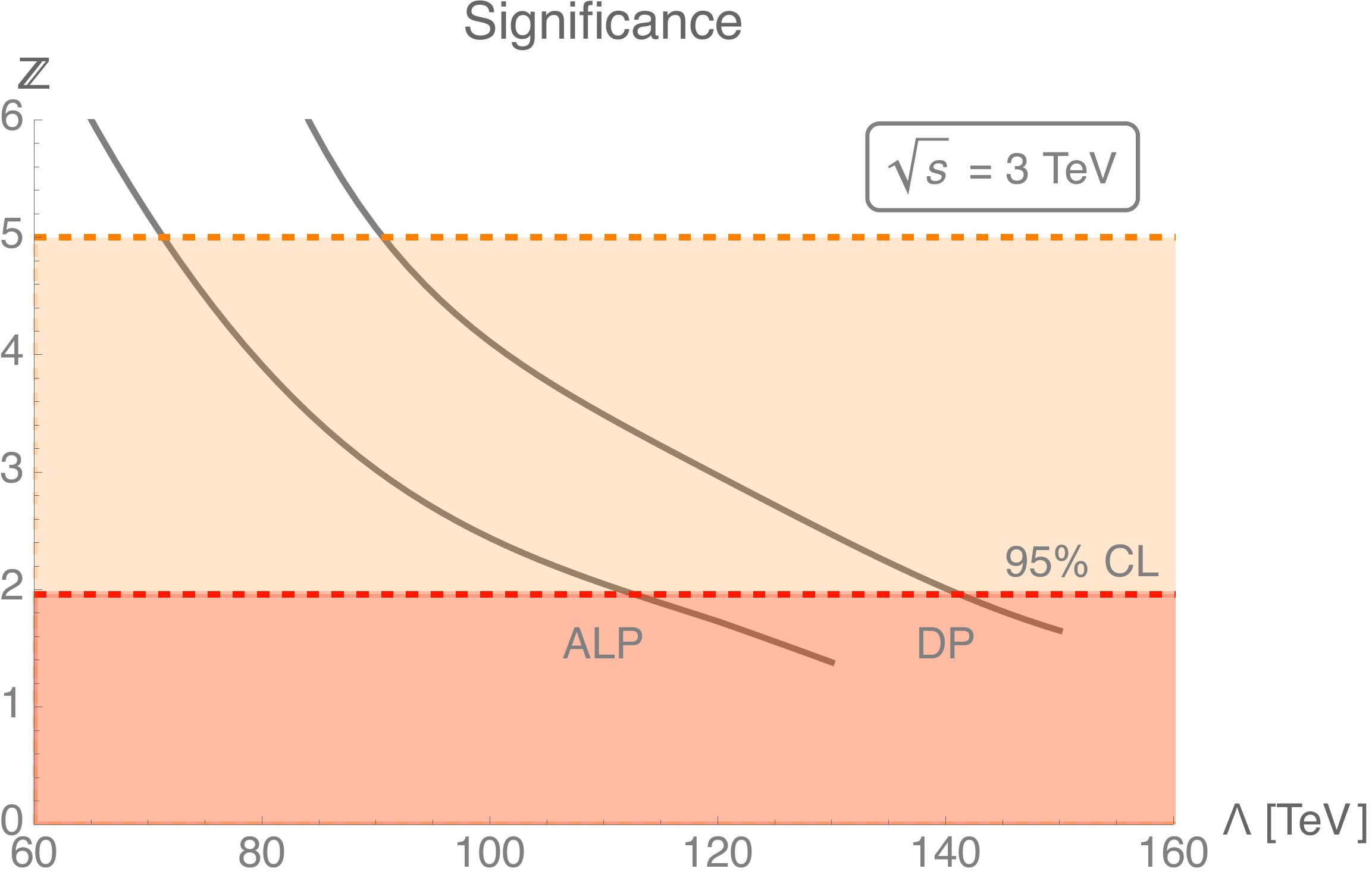}
\includegraphics[width=2.6in]{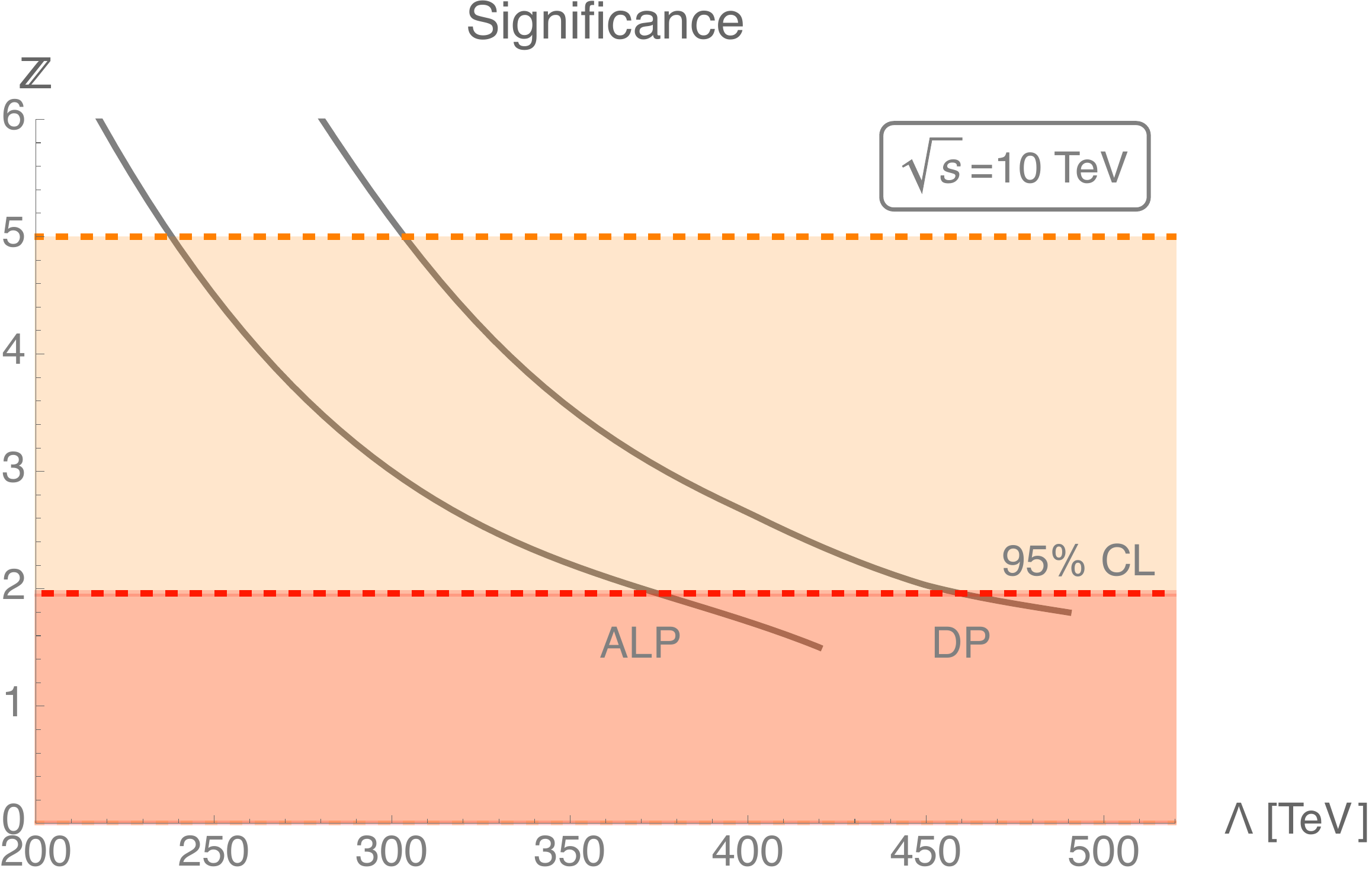}
\caption{\small 
\label{fig:significance} Significance for DP and ALP at $\sqrt{s}=3$ TeV (\underline{left}) and at $\sqrt{s}=10$ TeV (\underline{right}) as a function of the interaction scale $\Lambda$. Red dashed line corresponds to 95\% CL, orange dashed line to 5$\sigma$ (discovery reach).  The curves are obtained by running the simulation with a mass of 1 GeV for the DP or the ALP. The cross section is  independent of the mass of the DP or ALP, in the regime of several TeV we are exploring, as long as they are below 100 GeV.}
\end{center}
\end{figure*}

 We construct two statistical samples for the LLR, the first one with events characterized by the value of  $p^T_i$ generated according to the background-only population, the second one with $p^T_i$ generated according to the signal plus background population.
Fig.~\ref{LLR} shows the $p_T^\gamma$ distribution for signal and background events and the
result for the LLR analysis in the case of the dark photon at $\sqrt{s}=10$ TeV.

To quantify the difference in terms of statistical significance, we compute the p-value 
of the median of the signal-plus-background LLR distribution by integrating the background-only curve from the median value (indicated by the gray vertical line in Fig.~\ref{LLR})  to $+\infty$.  
The  significance is defined as ${\cal Z} = \Phi^{-1}(1-p)$
where
\be
\Phi(x) = \frac{1}{2}\, \left[ 1 + \text{erf} \left(\frac{ x}{\sqrt{2}} \right) \right]\,.
\ee
The value of  ${\cal Z}$ assigns a statistical significance to the separation between the two LLR distributions. We can take $ {\cal Z}$ as the number of $\sigma$'s, in the approximation in which the distribution is assumed to be Gaussian, and translates the number of $\sigma$'s into a confidence level (CL).

The significance thus obtained scales as $1/\Lambda^2$. We can plot the significance as a function of $\Lambda$ and find the effective scale value at which it is equal to 1.96. The value thus found corresponds to the largest value of $\Lambda$ for which we can separate the signal from the background with a CL of 95\%.

Figure~\ref{fig:significance} shows the determination of the scale $\Lambda$ for which the separation of the signal from the background reaches the 95\% CL for both the DP and the ALP at the two CM energies under consideration. Table~\ref{tab:lambda} gives the values of $\Lambda$ for the DP and ALP thus determined. The same table also gives the discovery (5$\sigma$)  for the largest $\Lambda$ reachable.
\begin{table}[h!]
\begin{tabular}{lcccc}
&\multicolumn{2}{c}{\hskip0.2cm $\sqrt{s}=3$ TeV \hskip0.2cm} &  \multicolumn{2}{c}{\hskip0.2cm $\sqrt{s}=10$ TeV \hskip0.2cm} \\[0.2cm]
\hline\\
& \color{persianblue}{DP} & \color{persianblue}{ALP} &\color{persianblue}{DP} & \color{persianblue}{ALP}  \\[0.2cm] %
 \underline{Limit} &\hskip0.2cm $141$ TeV \hskip0.2cm &\hskip0.2cm $112$ TeV\hskip0.2cm &\hskip0.2cm $459$ TeV \hskip0.2cm &\hskip0.2cm$375$ TeV 
  \\[0.2cm] %
\underline{Discovery} &\hskip0.2cm $92$ TeV \hskip0.2cm &\hskip0.2cm $71$ TeV\hskip0.2cm &\hskip0.2cm $303$ TeV \hskip0.2cm &\hskip0.2cm$238$ TeV\hskip0.2cm\\[0.4cm]
 \hline%
\end{tabular}
\caption{\label{tab:lambda} Explorable values of the effective energy scale $\Lambda$ for DP and ALP (95\% CL) for the two benchmark scenarios of the future muon collider under consideration.}
\end{table}

These results can be compared with current and future limits from cosmological, astrophysical and collider physics we discussed in the Introduction. In Fig.~\ref{bounds2} all the available limits for the coupling between the DP and muons are plotted together. A backward glance shows  that
\begin{itemize}
\item For massless DP the coupling is constrained by the SN events to a very small value, much smaller than those explorable at the muon collider; 
\item For massive DP and masses above 10 MeV the muon collider  could  provide new (and very stringent) limits. 
\end{itemize}

 The same is done for the limits for the coupling between the ALP and the photon in Fig.~\ref{bounds1}.   We see in this case that the muon collider could provide the best limits even though the estimated reach of Belle-II seems to overlap with that of the muon collider at  CM energy of 10 TeV.

\subsection{Distinguishing the DP from the ALP}

Assuming that a signal has been seen, will it be possible to determine whether it comes from a DP or an ALP? The difference rests on their characteristic angular distributions: while the DP angular distribution is flat, the ALP shows an angular dependence.

 \begin{figure*}[h!]
\begin{center}
\includegraphics[width=3in]{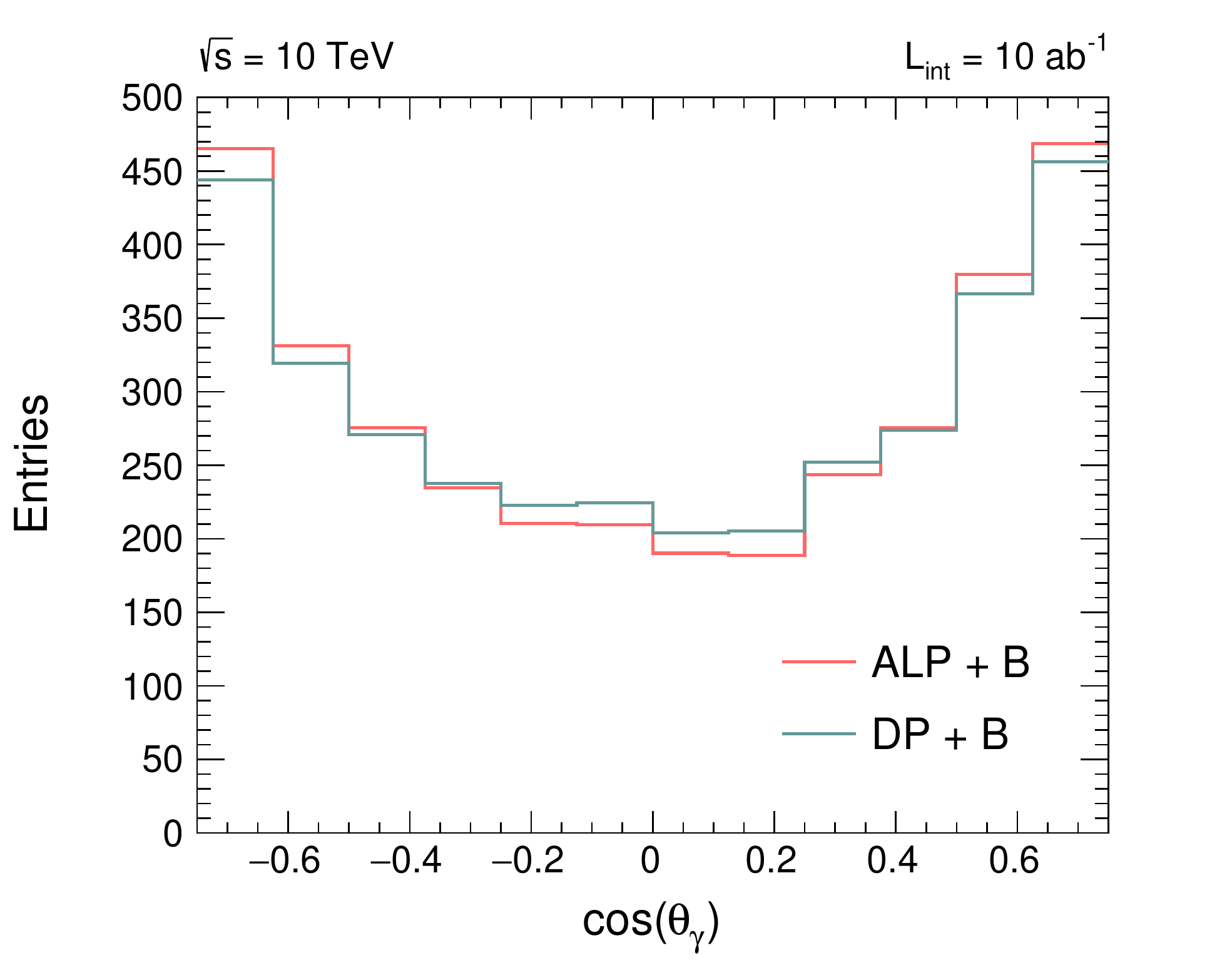}
\includegraphics[width=2.5in]{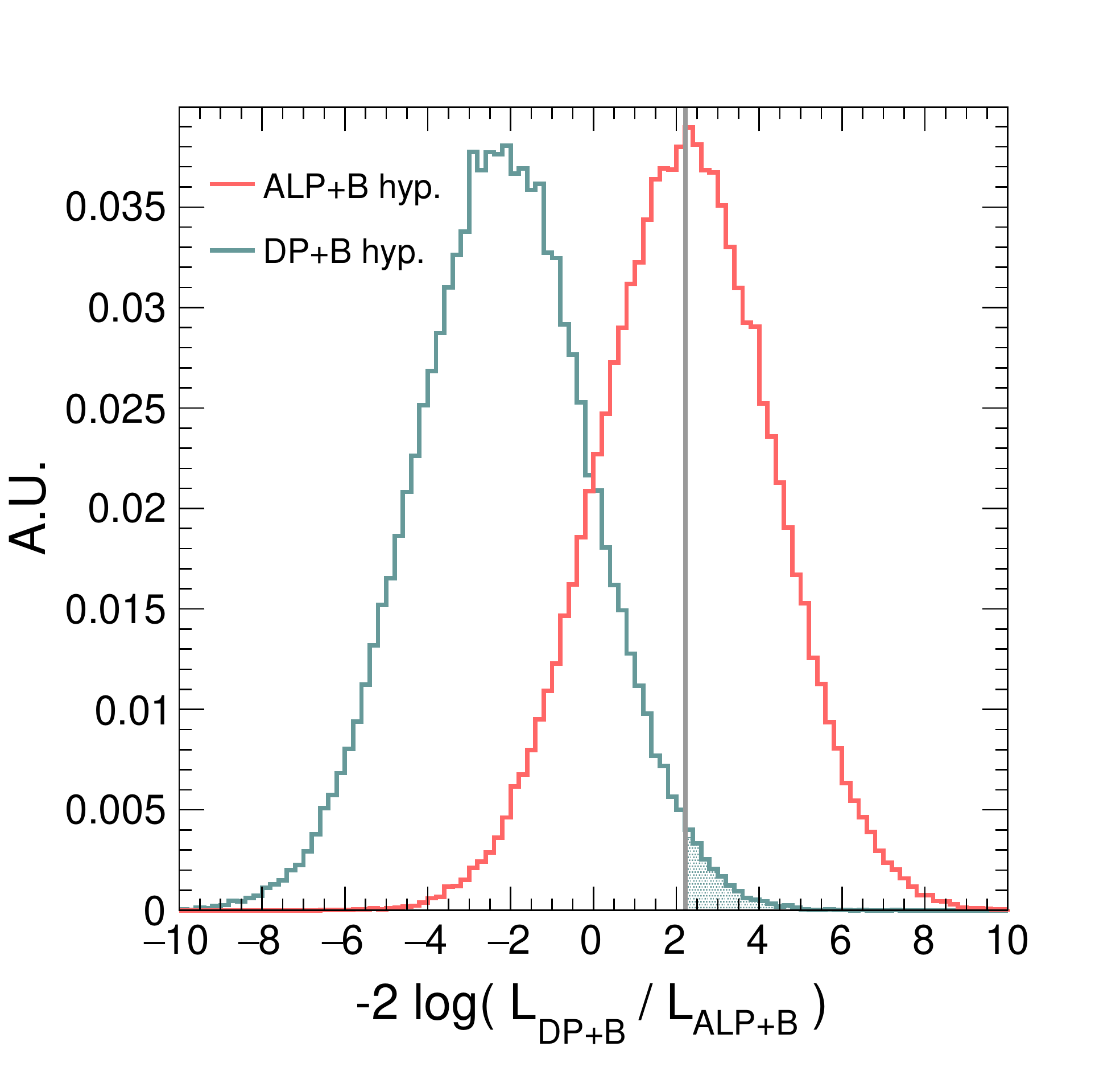}
\caption{\small 
\label{fig:angles}
\underline{On the left}: total angular distributions of the ALP and DP signals plus the background (B). 500 events are assumed for both signals. \underline{On the right}: LLR of the corresponding hypotheses.
 }
\end{center}
\end{figure*}

By means of a statistical analysis similar to that of the previous section,  we defined two likelihood functions, ${\cal L}_{\text{DP+B}}$ and  ${\cal L}_{\text{ALP+B}}$, including
angular pdf's for the signal and the background components in Eq.~\ref{eq:likelihood}.
The signal pdf's are parameterized as  functions of $\cos(\theta_\gamma)$ with a constant and a second order polynomial for the DP and the ALP hypotheses, respectively. The background pdf is a second order polynomial. Fig.~\ref{fig:angles} shows a comparison of the total angular distributions for the two cases and the corresponding LLR distributions.

We find that 500 events are necessary to distinguish with 95\% CL between the spin-0 (ALP) and the spin-1 (DP) hypotheses. For a scale $\Lambda\simeq 160$ TeV, this number of events is accumulated approximately in five years. A comparable number of events is necessary at $\sqrt{s}=3$ TeV.

\section{Conclusions}

The exploration of the physics program at a future muon collider has just begun. In this paper, we study the potential of the mono-photon signature in the search  for a dark sector. The high energy and luminosity made available by the muon collider make it the ideal machine to study those  interactions between the Standard Model and the dark sector from  dimension 5 portal operators  that grow with the energy, namely, those of the DP and the ALP.  Although  the sensitivity of the analysis is hampered by the radiative return of $Z$-boson  pole---which makes the cross section of the background non negligible at the end of the energy spectrum of the photons---we find that a muon collider could provide a competitive determination of the effective scale of the relevant operators.

A suitable choice of cuts on the photon energy and polar angle, to suppress the large  background induced by the radiative return effect, has been implemented to increase signal over background sensitivity.
We considered two benchmark scenarios corresponding to collision center-of-mass energies of
3 and 10 TeV, and  integrated (5 years) luminosities of 1 and 10 ab$^{-1}$, respectively.
In the case of non-observation of a signal, lower bounds at 95\% CL on the interaction scale $\Lambda$ for the dark-photon and ALP couplings have been derived for the 3 TeV collider corresponding to  $\Lambda=141$ and $\Lambda=112$ TeV respectively, that can be raised to $\Lambda=375$ TeV and $\Lambda=459$ TeV for 
the 10 TeV energy collisions respectively. A more sophisticated physical analysis based on two-dimensional cuts or multivariate techniques, which we leave to future work, may further improve the above sensitivities to the effective scale $\Lambda$.

When and if a signal is found, it will be important to know which dark sector particle is responsible for it.   We show that a muon collider operating at 3 or 10 TeV has  the potential to distinguish the spin-0 ALP from the spin-1 DP scenario. For a common energy scale $\Lambda=300$ TeV---about 500 events (which  can be approximatively accumulated in five years) are required to separate the two spin scenarios at the 95\% CL.

Our analysis  shows that a muon collider can provide a powerful tool in searching for  dark particles because of  its high sensitivity to the interactions scales of dark-boson portals, 
the clean environment of lepton collisions and the high luminosity.

\vskip2em
     {\small
\textit{Acknowledgements.---}
This work was performed within the Muon Collider Detector Design and Performance group~\cite{MCDDPG}. 
The digital inclusion of some of the experimental limits in the figures was done by means of \textsc{WebPlotDigitizer}~\cite{WebPlotDigitizer}.
MC thanks Federico Meloni for his comments and feedback.
  MF is affiliated to the Physics Department of the University of Trieste and SISSA---the support of which is acknowledged. 
MF and EG are affiliated to the Institute for Fundamental Physics of the Universe, Trieste, Italy.}

\vskip 2cm



\begin{thebibliography}{99}

\bibitem{Long:2020wfp}
K.~Long, D.~Lucchesi, M.~Palmer, N.~Pastrone, D.~Schulte and V.~Shiltsev,
``Muon colliders to expand frontiers of particle physics,''
Nature Phys. \textbf{17}, no.3, 289-292 (2021)
[\hhref{arXiv:2007.15684} [physics.acc-ph]].

\bibitem{Delahaye:2019omf}
J.~P.~Delahaye, M.~Diemoz, K.~Long, B.~Mansouli\'e, N.~Pastrone, L.~Rivkin, D.~Schulte, A.~Skrinsky and A.~Wulzer,
``Muon Colliders,''
[\hhref{arXiv:1901.06150} [physics.acc-ph]].


\bibitem{Greco:2016zli}
M.~Greco,
``Physics potential and motivations for a muon collider,''
Int. J. Mod. Phys. A \textbf{31} (2016) no.18, 1630028.


\bibitem{AlAli:2021let}
H.~Al Ali, N.~Arkani-Hamed, I.~Banta, S.~Benevedes, D.~Buttazzo, T.~Cai, J.~Cheng, T.~Cohen, N.~Craig and M.~Ekhterachian, \textit{et al.}
``The Muon Smasher's Guide,''
[\hhref{arXiv:2103.14043} [hep-ph]].

\bibitem{Barger:1995hr}
V.~D.~Barger, M.~S.~Berger, J.~F.~Gunion and T.~Han,
``Higgs Boson physics in the s channel at $\mu^+ \mu^-$ colliders,''
Phys. Rept. \textbf{286} (1997), 1-51
[\hhref{arXiv:hep-ph/9602415} [hep-ph]];\\
T.~Han and Z.~Liu,
``Potential precision of a direct measurement of the Higgs boson total width at a muon collider,''
Phys. Rev. D \textbf{87} (2013) no.3, 033007
[\hhref{arXiv:1210.7803} [hep-ph]];\\
A.~Conway and H.~Wenzel,
``Higgs Measurements at a Muon Collider,''
[\hhref{arXiv:1304.5270} [hep-ex]];\\
Y.~Alexahin, C.~M.~Ankenbrandt, D.~B.~Cline, A.~Conway, M.~A.~Cummings, V.~Di Benedetto, E.~Eichten, C.~Gatto, B.~Grinstein and J.~Gunion, \textit{et al.}
``Muon Collider Higgs Factory for Snowmass 2013,''
[\hhref{arXiv:1308.2143} [hep-ph]];\\
G.~J.~Gounaris and F.~M.~Renard,
``Test of the triple Higgs boson form factor in $\mu^-\mu^+\to HH$,''
Phys. Rev. D \textbf{93} (2016), 093018
[\hhref{arXiv:1601.04142} [hep-ph]];\\
T.~Han, D.~Liu, I.~Low and X.~Wang,
``Electroweak couplings of the Higgs boson at a multi-TeV muon collider,''
Phys. Rev. D \textbf{103}, no.1, 013002 (2021)
[\hhref{arXiv:2008.12204} [hep-ph]];\\
M.~Chiesa, F.~Maltoni, L.~Mantani, B.~Mele, F.~Piccinini and X.~Zhao,
``Measuring the quartic Higgs self-coupling at a multi-TeV muon collider,''
JHEP \textbf{09} (2020), 098
[\hhref{arXiv:2003.13628} [hep-ph]].



\bibitem{Buttazzo:2020uzc}
D.~Buttazzo, R.~Franceschini and A.~Wulzer,
``Two Paths Towards Precision at a Very High Energy Lepton Collider,''
JHEP \textbf{05} (2021), 219
[\hhref{arXiv:2012.11555} [hep-ph]].

\bibitem{Han:2020pif}
T.~Han, D.~Liu, I.~Low and X.~Wang,
``Electroweak couplings of the Higgs boson at a multi-TeV muon collider,''
Phys. Rev. D \textbf{103}, no.1, 013002 (2021)
[\hhref{arXiv:2008.12204} [hep-ph]].

\bibitem{Costantini:2020stv}
A.~Costantini, F.~De Lillo, F.~Maltoni, L.~Mantani, O.~Mattelaer, R.~Ruiz and X.~Zhao,
``Vector boson fusion at multi-TeV muon colliders,''
JHEP \textbf{09} (2020), 080
[\hhref{arXiv:2005.10289} [hep-ph]].



\bibitem{Asadi:2021gah}
P.~Asadi, R.~Capdevilla, C.~Cesarotti and S.~Homiller,
``Searching for leptoquarks at future muon colliders,''
JHEP \textbf{10}, 182 (2021)
[\hhref{arXiv:2104.05720} [hep-ph]];\\
S.~Qian, C.~Li, Q.~Li, F.~Meng, J.~Xiao, T.~Yang, M.~Lu and Z.~You,
``Searching for heavy leptoquarks at a muon collider,''
JHEP \textbf{12}, 047 (2021)
[\hhref{arXiv:2109.01265} [hep-ph]].

\bibitem{Han:2020uak}
T.~Han, Z.~Liu, L.~T.~Wang and X.~Wang,
``WIMPs at High Energy Muon Colliders,''
Phys. Rev. D \textbf{103}, no.7, 075004 (2021)
[\hhref{arXiv:2009.11287} [hep-ph]];\\
R.~Capdevilla, F.~Meloni, R.~Simoniello and J.~Zurita,
``Hunting wino and higgsino dark matter at the muon collider with disappearing tracks,''
JHEP \textbf{06}, 133 (2021)
[\hhref{arXiv:2102.11292} [hep-ph]].

 \bibitem{dark_sector} 
  R.~Essig {\it et al.},
  ``Working Group Report: New Light Weakly Coupled Particles,''
  \hhref{arXiv:1311.0029} [hep-ph];
  J.~Alexander {\it et al.},
  ``Dark Sectors 2016 Workshop: Community Report,''
 [\hhref{arXiv:1608.08632} [hep-ph]];
  M.~A.~Deliyergiyev,
  ``Recent Progress in Search for Dark Sector Signatures,''
  Open Phys.\  {\bf 14}, no. 1, 281 (2016)
  [\hhref{arXiv:1510.06927} [hep-ph]].
  

\bibitem{Holdom:1985ag}
B.~Holdom,
``Two U(1)'s and Epsilon Charge Shifts,''
Phys. Lett. B \textbf{166}, 196-198 (1986)



\bibitem{Fabbrichesi:2020wbt}
M.~Fabbrichesi, E.~Gabrielli and G.~Lanfranchi,
\href{https://doi.org/10.1007/978-3-030-62519-1}{\textit{The Physics of the Dark Photon. A Primer}}, SpringerBriefs in Physics  (2020) [\hhref{arXiv:2005.01515} [hep-ph]].

\bibitem{Dobrescu:2004wz}
B.~A.~Dobrescu,
``Massless gauge bosons other than the photon,''
Phys. Rev. Lett. \textbf{94}, 151802 (2005)
[\hhref{arXiv:hep-ph/0411004 }[hep-ph]].


\bibitem{Gabrielli:2016cut}
E.~Gabrielli, B.~Mele, M.~Raidal and E.~Venturini,
``FCNC decays of standard model fermions into a dark photon,''
Phys. Rev. D \textbf{94}, no.11, 115013 (2016)
[\hhref{arXiv:1607.05928} [hep-ph]].



  
\bibitem{Peccei:1977hh}
R.~D.~Peccei and H.~R.~Quinn,
``CP Conservation in the Presence of Instantons,''
Phys. Rev. Lett. \textbf{38}, 1440-1443 (1977)

\bibitem{Weinberg:1977ma}
S.~Weinberg,
``A New Light Boson?,''
Phys. Rev. Lett. \textbf{40}, 223-226 (1978)


\bibitem{Wilczek:1977pj}
F.~Wilczek,
``Problem of Strong  $P$  and  $T$  Invariance in the Presence of Instantons,''
Phys. Rev. Lett. \textbf{40}, 279-282 (1978)

\bibitem{Bento:1985in}
L.~Bento, J.~C.~Romao and A.~Barroso,
``$e^+ e^- \to \gamma$ + missing neutrals: neutrinos vs.\ photino production,''
Phys. Rev. D \textbf{33} (1986), 1488

\bibitem{Berends:1987zz}
F.~A.~Berends, G.~J.~H.~Burgers, C.~Mana, M.~Martinez and W.~L.~van Neerven,
``Radiative Corrections to the Process $e^+ e^- \to$ Neutrino Anti-neutrino $\gamma$,''
Nucl. Phys. B \textbf{301} (1988), 583-600.

\bibitem{Chakrabarty:2014pja}
N.~Chakrabarty, T.~Han, Z.~Liu and B.~Mukhopadhyaya,
``Radiative Return for Heavy Higgs Boson at a Muon Collider,''
Phys. Rev. D \textbf{91} (2015) no.1, 015008
[\hhref{arXiv:1408.5912} [hep-ph]].


\bibitem{Mimasu:2014nea}
K.~Mimasu and V.~Sanz,
``ALPs at Colliders,''
JHEP \textbf{06} (2015), 173
[\hhref{arXiv:1409.4792} [hep-ph]].

 
\bibitem{Bauer:2018uxu}
M.~Bauer, M.~Heiles, M.~Neubert and A.~Thamm,
``Axion-Like Particles at Future Colliders,''
Eur. Phys. J. C \textbf{79} (2019) no.1, 74
[\hhref{arXiv:1808.10323} [hep-ph]].

\bibitem{Biswas:2019lcp}
S.~Biswas, A.~Chatterjee, E.~Gabrielli and B.~Mele,
``Probing dark-axion like particle portals at future $e^+e^-$ colliders,''
Phys. Rev. D \textbf{100} (2019) no.11, 115040
[\hhref{arXiv:1906.10608} [hep-ph]].


\bibitem{Darme:2020sjf}
L.~Darm\'e, F.~Giacchino, E.~Nardi and M.~Raggi,
``Invisible decays of axion-like particles: constraints and prospects,''
JHEP \textbf{06} (2021), 009
[\hhref{arXiv:2012.07894} [hep-ph]].


\bibitem{OPAL:1998aqw}
G.~Abbiendi \textit{et al.} [OPAL],
``Search for anomalous photonic events with missing energy in $e^+ e^-$ collisions at $\sqrt{s}= 130, 136$ and $183$ GeV,''
Eur. Phys. J. C \textbf{8} (1999), 23-40
[\hhref{arXiv:hep-ex/9810021} [hep-ex]].

\bibitem{L3:2003yon}
P.~Achard \textit{et al.} [L3],
``Single photon and multiphoton events with missing energy in $e^{+} e^{-}$ collisions at LEP,''
Phys. Lett. B \textbf{587} (2004), 16-32
[\hhref{arXiv:hep-ex/0402002} [hep-ex]].

\bibitem{DELPHI:2003dlq}
J.~Abdallah \textit{et al.} [DELPHI],
``Photon events with missing energy in e+ e- collisions at s**(1/2) = 130-GeV to 209-GeV,''
Eur. Phys. J. C \textbf{38} (2005), 395-411
[\hhref{arXiv:hep-ex/0406019} [hep-ex]].

\bibitem{CDF:2008njt}
T.~Aaltonen \textit{et al.} [CDF],
``Search for large extra dimensions in final states containing one photon or jet and large missing transverse energy produced in $p \bar{p}$ collisions at $\sqrt{s}$ = 1.96-TeV,''
Phys. Rev. Lett. \textbf{101}, 181602 (2008)
[\hhref{arXiv:0807.3132} [hep-ex]].

\bibitem{D0:2008ayi}
V.~M.~Abazov \textit{et al.} [D0],
``Search for large extra dimensions via single photon plus missing energy final states at $\sqrt{s}$ = 1.96-TeV,''
Phys. Rev. Lett. \textbf{101}, 011601 (2008)
[\hhref{arXiv:0803.2137} [hep-ex]].

\bibitem{ATLAS:2016zxj}
M.~Aaboud \textit{et al.} [ATLAS],
``Search for new phenomena in events with a photon and missing transverse momentum in $pp$ collisions at $\sqrt{s}=13$ TeV with the ATLAS detector,''
JHEP \textbf{06} (2016), 059
[\hhref{arXiv:1604.01306} [hep-ex]].

\bibitem{CMS:2018ffd}
A.~M.~Sirunyan \textit{et al.} [CMS],
``Search for new physics in final states with a single photon and missing transverse momentum in proton-proton collisions at $\sqrt{s} =$ 13 TeV,''
JHEP \textbf{02} (2019), 074
[\hhref{arXiv:1810.00196} [hep-ex]].

 

\bibitem{Ringwald:2014vqa}
A.~Ringwald,
``Axions and Axion-Like Particles,''
[\hhref{arXiv:1407.0546} [hep-ph]].



\bibitem{Bollig:2020xdr}
R.~Bollig, W.~DeRocco, P.~W.~Graham and H.~T.~Janka,
``Muons in Supernovae: Implications for the Axion-Muon Coupling,''
Phys. Rev. Lett. \textbf{125}, no.5, 051104 (2020)
[erratum: Phys. Rev. Lett. \textbf{126}, no.18, 189901 (2021)]
[\hhref{arXiv:2005.07141} [hep-ph]].





\bibitem{DEramo:2018vss}
F.~D'Eramo, R.~Z.~Ferreira, A.~Notari and J.~L.~Bernal,
``Hot Axions and the $H_0$ tension,''
JCAP \textbf{11} (2018), 014
[\hhref{arXiv:1808.07430} [hep-ph]].


\bibitem{Pospelov:2008zw}
M.~Pospelov,
``Secluded U(1) below the weak scale,''
Phys. Rev. D \textbf{80} (2009), 095002
[\hhref{arXiv:0811.1030} [hep-ph]].


\bibitem{Escudero:2019gzq}
M.~Escudero, D.~Hooper, G.~Krnjaic and M.~Pierre,
``Cosmology with A Very Light L$_{\mu}$ \ensuremath{-} L$_{\tau}$ Gauge Boson,''
JHEP \textbf{03} (2019), 071
[\hhref{arXiv:1901.02010} [hep-ph]].

\bibitem{NA64:2020qwq}
D.~Banerjee \textit{et al.} [NA64],
``Search for Axionlike and Scalar Particles with the NA64 Experiment,''
Phys. Rev. Lett. \textbf{125} (2020) no.8, 081801
[\hhref{arXiv:2005.02710} [hep-ex]].

\bibitem{DELPHI:2008uka}
J.~Abdallah \textit{et al.} [DELPHI],
``Search for one large extra dimension with the DELPHI detector at LEP,''
Eur. Phys. J. C \textbf{60} (2009), 17-23
[\hhref{arXiv:0901.4486} [hep-ex]].

\bibitem{BaBar:2017tiz}
J.~P.~Lees \textit{et al.} [BaBar],
``Search for Invisible Decays of a Dark Photon Produced in ${e}^{+}{e}^{-}$ Collisions at BaBar,''
Phys. Rev. Lett. \textbf{119} (2017) no.13, 131804
[\hhref{arXiv:1702.03327} [hep-ex]].




\bibitem{Belle-II:2018jsg}
E.~Kou \textit{et al.} [Belle-II],
``The Belle II Physics Book,''
PTEP \textbf{2019} (2019) no.12, 123C01
[erratum: PTEP \textbf{2020} (2020) no.2, 029201]
[\hhref{arXiv:1808.10567} [hep-ex]].

\bibitem{Dolan:2017osp}
M.~J.~Dolan, T.~Ferber, C.~Hearty, F.~Kahlhoefer and K.~Schmidt-Hoberg,
``Revised constraints and Belle II sensitivity for visible and invisible axion-like particles,''
JHEP \textbf{12} (2017), 094
[erratum: JHEP \textbf{03} (2021), 190]
[\hhref{arXiv:1709.00009} [hep-ph]].


\bibitem{Bjorken:1988as}
J.~D.~Bjorken, S.~Ecklund, W.~R.~Nelson, A.~Abashian, C.~Church, B.~Lu, L.~W.~Mo, T.~A.~Nunamaker and P.~Rassmann,
``Search for Neutral Metastable Penetrating Particles Produced in the SLAC Beam Dump,''
Phys. Rev. D \textbf{38} (1988), 3375.

\bibitem{BIB}
N.~Bartosik \etal, ``Preliminary Report on the Study of Beam-Induced Background Effects at a Muon Collider,''  \hhref{arXiv:1905.03725v1};
F.~Collamati {\it et al.}, ``Advanced assessment of Beam Induced Background at a Muon Collider,''
[\hhref{arXiv:2105.09116}].

\bibitem{MuColDet1} N.~Bartosik \etal, ``Detector and Physics Performance at a Muon Collider,''
\href{https://doi.org/10.1088/1748-0221/15/05/p05001}{2020 JINST {\bf 15}, P05001 (2020)}. 

\bibitem{MuColDet2} N.~Bartosik \etal, ``Full Detector Simulation with Unprecedented Background Occupancy at a Muon Collider,'' 
\href{https://doi.org/10.1007/s41781-021-00067-x}{Comput Softw Big Sci \textbf{5} (2021) 21}. 

\bibitem{MG5}
J.~Alwall \etal,
``The automated computation of tree-level and next-to-leading order differential cross sections, and their matching to parton shower simulations,''
JHEP \textbf{07} (2014), 079
[\hhref{arXiv:1405.0301} [hep-ph]].

\bibitem{Sjostrand:2006za}
T.~Sjostrand, S.~Mrenna and P.~Z.~Skands,
``PYTHIA 6.4 Physics and Manual,''
JHEP \textbf{05} (2006), 026
[\hhref{arXiv:hep-ph/0603175} [hep-ph]].

\bibitem{Sjostrand:2014zea}
T.~Sj\"ostrand, S.~Ask, J.~R.~Christiansen, R.~Corke, N.~Desai, P.~Ilten, S.~Mrenna, S.~Prestel, C.~O.~Rasmussen and P.~Z.~Skands,
``An introduction to PYTHIA 8.2,''
Comput. Phys. Commun. \textbf{191} (2015), 159-177
[\hhref{arXiv:1410.3012} [hep-ph]].

\bibitem{ILCSoft}
 M.~Frank {\it et al.}, ``DD4hep: A Detector Description Toolkit for High Energy Physics Experiments,''
  \href{https://doi.org/10.1088/1742-6596/513/2/022010}{2014 J. Phys.: Conf. Ser. {\bf 513} 022010};
  F.~Gaede {\it et al.}, ``LCIO: A Persistency framework for linear collider simulation studies,''
  CHEP-2003-TUKT001 [\hhref{arXiv:physics/0306114v1}];
  F.~Gaede, ``Marlin and LCCD--Software tools for the ILC,''
  \href{https://doi.org/10.1016/j.nima.2005.11.138}{Nucl. Inst. and Methods Phys. Res. A \textbf{559} (2006) 177-180}.
  
\bibitem{CLICdet}
``Detector Technologies for CLIC,'' edited by D.~Dannheim, K.~Kr\"uger, A.~Levy, 
A.~N\"urnberg, E.~Sicking,
\href{http://dx.doi.org/10.23731/CYRM-2019-001}{CERN–2019–001} (CERN, Geneva, 2019)
[\hhref{arXiv:1905.02520v1}].

\bibitem{Pandora}
J.S.~Marshall and M.A.~Thomson, ``The Pandora software development kit for pattern recognition,''
\href{https://doi.org/10.1140/epjc/s10052-015-3659-3}{Eur. Phys. J. C \textbf{75} (2015) 439}.

\bibitem{MCDDPG}
The Muon Collider Detector Design and Performance group,
\href{https://muoncollider.web.cern.ch/design/muon-collider-detector}{ https://muoncollider.web.cern.ch/design/muon-collider-detector}.

\bibitem{WebPlotDigitizer}
\url{https://automeris.io/WebPlotDigitizer}.
\end{thebibliography}
\end{document}